\documentclass[1p]{elsarticle}

\usepackage{lineno,hyperref}
\usepackage{subfigure}
\usepackage{amssymb}
\modulolinenumbers[5]

\journal{Annals of Physics}

\begin{document}

\begin{frontmatter}

\title{Renormalization Group Flow of the Aharonov-Bohm Scattering Amplitude}

%% Group authors per affiliation:
\author{U. Camara da Silva\fnref{myfootnote}}
\address{Departamento de F\'\i sica - CCE\\
Universidade Federal de Espirito Santo\\
29075-900, Vitoria - ES, Brazil}
\fntext[myfootnote]{ulyssescamara@gmail.com}

%% or include affiliations in footnotes:
%\author[Departamento de F\'\i sica - CCE\\
%Universidade Federal de Espirito Santo\\
%29075-900, Vitoria - ES, Brazil]{Elsevier Inc}
%\ead[url]{www.elsevier.com}

%\author[mysecondaryaddress]{Global Customer Service\corref{mycorrespondingauthor}}
%\cortext[mycorrespondingauthor]{Corresponding author}
%\ead{support@elsevier.com}

%\address[mymainaddress]{1600 John F Kennedy Boulevard, Philadelphia}
%\address[mysecondaryaddress]{360 Park Avenue South, New York}

%\tableofcontents

\begin{abstract}
The Aharonov-Bohm elastic scattering with incident particles described by plane waves is revisited  by using the phase-shifts method. The formal equivalence between the cylindrical Schr\"odinger equation and the one-dimensional Calogero problem allows us to show that up to two scattering phase-shifts modes in the cylindrical waves expansion must be renormalized. The renormalization procedure introduces new length scales giving rise to spontaneous breaking of the conformal symmetry. The new renormalized cross-section has an amazing property of being non-vanishing  even for a quantized magnetic flux, coinciding with the case of Dirac delta function potential. The knowledge of the exact beta function permits us to describe the renormalization group flows  within the  two-parametric family of  renormalized  Aharonov-Bohm scattering amplitudes. Our analysis demonstrates that for quantized magnetic fluxes a  BKT-like phase transition at the coupling space occurs.
\end{abstract}


\end{frontmatter}

%\linenumbers

\section{Introduction}

In the presence of an external electromagnetic field, the wave function of a charged particle in quantum mechanics contains a non-integrable phase factor, which is responsible for observable quantum effects in the case of  non-simply connected spaces. The most representative example of such  phenomenon is the famous Aharonov-Bohm\footnote{ Extensive reviews on the subject can be  found in refs. \cite{Peshkin,Tonomura_new}.} (AB) magnetic effect \cite{AB_original}, experimentally confirmed  in ref. \cite{experimento_1_AB}. It has  provided also important insights  for  generalizations to the case of  non-abelian gauge fields, where the analogous phase factor turns out  to play a relevant role even for simply-connected spaces \cite{Wu_Yang}.

The present article is dedicated to the rigorous study of two long-standing problems of the original AB elastic scattering for non-relativistic spinless incident particle with well defined momentum, i.e. the plane wave. Since  the minimal coupling introduces  a non-integrable phase in the incident plane wave, related to the magnetic flux $\Phi$, the usual scattering boundary condition cannot be  trivially implemented, due to  the apparent multivaluedness\footnote{See ref. \cite{Merzbacher} for  a comprehensible discussion concerning the  single-valuedness condition for the spinless particle  wave function.} of the wave function. In the first part of the paper we review the exact phase-shift method, where the problematic phase is transferred to the outgoing cylindrical wave, that allows to derive the  AB scattering amplitude in a rather simple way \cite{Ruji_AB}. This approach however  introduces  an unexpected   ``side effect'', namely  an  extra $\delta$-function term in the scattering amplitude, that will be discussed in Sect. \ref{sec:esp}. Within the framework of this method, the total wave function is realized as a superposition of cylindrical waves of different angular momenta $m=0,\pm1,\pm2 ,...$, each one of them obeying a Dirichlet boundary condition (b.c.) at the origin. The development of  modern mathematical tools for  self-adjoint (s.a.) extensions of second order differential operators \cite{Gitman}, as the one given by the cylindrical AB Hamiltonian $H_{cal}= - \frac{d^2}{dr^2}+V(r)$, pointed out that  the above statement fails to be valid for certain range of  values of the effective coupling constant $\mathcal C_m=(m+\alpha)^2< \frac{3}{4}$  of the  Calogero potential  $V(r)=\mathcal C_m/r^2$. The correct implementation of the s.a. condition for the  AB  operator  $H_{cal}$ requires the consideration of a one-parametric family of more general Dirichlet  or mixed  b.c.'s for the corresponding AB cylindrical wave functions. Although this problem and its mathematical solution have already been widely discussed in the literature \cite{Gitman,Beane,Gitman2}, their implications  for  AB scattering amplitudes are not yet  completely explored and far from being  fully understood.

The second part of the paper represents an attempt for a complete description of all the physical consequences coming from the correct  s.a. extension of the cylindrical AB Hamiltonian. In section \ref{sec:ren}, by taking into account the correspondence of the AB cylindrical equation with the one dimensional Calogero problem \cite{Cal1,Cal3}, we solve the b.c. problem  by using the renormalization method of Beane \emph{et al.} \cite{Beane}. As a result  of this renormalization the scattering amplitude acquires up to two new scale parameters, characterizing the conformal symmetry breaking and thus providing  an example of dimensional transmutation \cite{Wein_Col}. It  also introduces  certain important corrections to  the original AB cross-section. The most impressive result of this renormalization procedure  is the prediction of the explicit form of a nontrivial isotropic cross-section in the case of ``quantized''\footnote{i.e. an integer number of the fundamental flux $\Phi_0=2\pi\hbar/|e|$, where $e$ denotes the electric  charge  of the particle.} magnetic flux $\Phi=n\Phi_0$. In the final part of section \ref{sec:ren} the conditions for the existence of a few bound states consistent with the AB scattering problem are explored. They can be considered as the remnants of the  Landau levels that have survived after the zero radius limit of the AB solenoid. The exact beta function of the running coupling and related RG flows are studied in details in section \ref{sec:beta}. Again, the case of quantized magnetic flux exhibits  particularly interesting features:  only one mode  needs to be renormalized and the two RG fixed points in the coupling space  collide  into a BKT-like critical point of an  infinite order phase transition \cite{B1_BKT,B2_BKT,KT_BKT}. A discussion about the physical meaning of the new quantum scales is presented at our concluding section 5. 
 
\section{AB scattering on infinitely thin solenoid }\label{sec:esp}
The original AB effect concerns a scattering process involving a charged particle and an infinite ideal solenoid. The solenoid is placed along the $z$-axis and has a small radius $r_c$ and a strong magnetic field $\vec B=B\hat z$. The limits $r_c\rightarrow0$ and $B\rightarrow\infty$ are taken by keeping the magnetic flux $\Phi=\pi r_c^2B$  fixed and finite.  We will use the term ``infinitely thin solenoid'' for this configuration \cite{Gitman}. In the plane ortogonal  to the solenoid there is no magnetic field at any point, except for the origin of the coordinate system. The standard gauge choice given by  $\vec A=\frac{\Phi}{2\pi}\frac{\hat\varphi}{r}$, $r=\sqrt{x^2+y^2}\neq0$ provides $\oint_C\vec A.d\vec l=\Phi$ and $\vec B=\vec\nabla\times\vec A=\Phi\delta(x)\delta(y)\hat z$. Then the  stationary Schr\"odinger equation describing a spinless particle with mass $M$ and electric charge $-|e|$ interacting with  the infinitely thin solenoid takes the form:
\begin{eqnarray}
-\frac{\hbar^2}{2M}\left(\vec \nabla+i\frac{|e|}{\hbar}\vec A\right)^2\psi=E\psi.
\end{eqnarray}
We consider the general solution for zero momentum $p_z=0$,\footnote{So for all purposes we address a two-dimensional problem.} that can be decomposed  into a superposition of cylindrical waves in the $x-y$ plane:
\begin{eqnarray}
\phi(r,\varphi)=\sum_{m=-\infty}^{\infty}R_m(r)e^{im\varphi}, \ r=\sqrt{x^2+y^2},\  \tan\varphi=y/x.
\end{eqnarray}
The functions $R_m(r)$ are solutions of the cylindrical Schr\"odinger equation
\begin{eqnarray}
r^2R''_m(r)+rR'_m(r)+\left[(kr)^2-\left(m+\frac{\Phi}{\Phi_0}\right)^2\right]R_m(r)=0,\label{R}
\end{eqnarray}
which can be solved in terms of Bessel functions \cite{Bessel}
\begin{eqnarray}
R_m(r)=A_mJ_{\nu_m}(kr)+B_mN_{\nu_m}(kr), \ \nu_m=\left|m-\alpha\right|,\label{Be}
\end{eqnarray}
where $\alpha=\Phi/\Phi_0$, $\Phi_0\equiv 2\pi\hbar/|e|$.

The gradient form of the vector potential  $\vec A=\vec\nabla\omega(\varphi)$, $\omega(\varphi)=-\alpha\frac{\Phi_0}{2\pi}(\pi-\varphi)$, indicates that apparently  it represents a pure gauge field configuration. This assertion is not however true since the phase factor $e^{-i\frac{|e|}{\hbar}\omega(\varphi)}$ is a multiple-valued function,\footnote{In quantum mechanics the single-valuedness of the wave function does not require the continuity of the gauge transformation; the continuity of $\exp{(-i|e|\omega(\varphi)/\hbar)}$ is sufficient.} $\omega(2\pi)-\omega(0)=\alpha\Phi_0$, if $\alpha\Phi_0/2\pi\notin \, \mathbb{N}$. Notice that the free particle wave function also shares the same property due to the minimal coupling with the vector potential. Then
\begin{eqnarray}
\psi_{in}^{AB}(r,\varphi)=e^{ikx+i\alpha\left(\pi-\varphi\right)}, \ x=r\cos\varphi,\label{free_AB}
\end{eqnarray}
has a branch cut along  the positive $x$-axis, $\psi^{AB}_{in}(r,2\pi)/\psi^{AB}_{in}(r,0)=e^{-2i\pi\alpha}$, excepts the special case  $\alpha\in\mathbb{N}$ when the magnetic flux has discrete values.

\subsection{On the choice of scattering boundary conditions}\label{subsec:scattbcs}

According to the prescription of  the original  AB  paper, the function $\psi^{AB}_{in}(r,\varphi)$ must be given by the asymptotic form of the incident wave function, in order to ensure that the current density
\begin{eqnarray}
\vec j_{in}^{AB}=-i\frac{\hbar}{2M}\left[(\psi_{in}^{AB})^*\left(\vec\nabla-i\frac{e}{\hbar}\vec A\right)\psi_{in}^{AB}-\psi_{in}^{AB}\left(\vec\nabla+i\frac{e}{\hbar}\vec A\right)(\psi_{in}^{AB})^*\right]=\frac{\hbar k}{M}\hat x,
\end{eqnarray}
 is a  constant along the $x$ direction. Notice that  the b.c.
\begin{eqnarray}
\psi^{AB}(r,\phi)\approx e^{ikx+i\alpha\left(\pi-\varphi\right)}+f_k^{AB}(\varphi)\frac{e^{ikr}}{\sqrt{-ir}}, \ kr\gg1,\label{assin_AB}
\end{eqnarray}
 imposed in the original description of the AB scattering, \emph{does not} correspond to the standard asymptotic form  for a short-range interaction elastic scattering, as  explained  in \ref{sec:esp_2d} (see Eq. (\ref{1})).  It is usually argued \cite{AB_original} that the multi-valuedness of the incident wave function is expected to be canceled by an equivalent hidden term in the second function representing the cylindrical wave. Aharanov and Bohm have proposed a single-valued \emph{ansatz}, but the asymptotic expression they have obtained, due to the approximations performed during the calculations, is given by an exponential multi-valued function with a branch-cut along the  $x>0$ semi-axis. Only in a few  special cases, viz. for $\alpha=n$ and $\alpha=n+1/2$, $n\in\mathbb{N}$, the authors obtain asymptotically single-valued forms.

\vspace{0,5cm}

Our approach to the AB problem is based on a slight modification of the original scattering \emph{ansatz}, by transferring the extra phase in $\psi_{in}^{AB}$ from the incident wave to the outgoing cylindrical one. The exponential (\ref{free_AB}) describes a particle which is not interacting with the magnetic field, so locally we have a  pure gauge phase. However the fact that $e^{i\alpha(\pi-\varphi)}$ is multi-valued ($\alpha\notin\mathbb{N}$) points out  that such a description cannot be adopted over the whole plane. In the scattering process, the incident particle is released from $r\rightarrow\infty$ with $\varphi=\pi$, $\psi_{in}=\psi_{in}^{AB}(x,0)=e^{ikx}$, and goes to the $r\rightarrow\infty$ with $\varphi\rightarrow0$ or $2\pi$ after scattering. During the scattering, the particle  interacts with the solenoid's magnetic field, so that its wave function cannot be written in the form  $\psi_{in}^{AB}$ anymore. Now observe that the AB incident wave can be rewritten as 
\begin{eqnarray}
\psi_{in}^{AB}=e^{ikx+i\alpha\left(\pi-\varphi\right)}=e^{ikx}+2 ie^{i\frac{\alpha}{2}\left(\pi-\varphi\right)}\sin\left(\frac{\alpha}{2}\left(\pi-\varphi\right)\right)e^{ikx},\label{AB_in}
\end{eqnarray}
thus separating the standard incident wave $e^{ikx}$ from the multi-valued term. Since we are dealing with an elastic scattering from a short range interaction, the scattered wave function  $\psi_{scat}$,  by definition  corresponds to the difference $\psi(r,\varphi)-\psi_{in}(r,\varphi)$, with $r\rightarrow\infty$, $\forall$ $\varphi$. It is therefore  natural to interpret the second term of (\ref{AB_in}) as a part of $\psi_{scat}$. Then assuming that $\psi_{in}=e^{ikx}$, the proper scattering b.c. is now given by equation (\ref{1}). There is no inconsistency with the probability current density, since the $\vec j_{in}=\frac{\hbar k}{M}\hat x+\mathcal{O}(1/r)$ continues to be a constant in the limit $r\rightarrow\infty$.
%%%%%%%%%%%%%%%%%%%%%%%%%%%%%%%%%%%%%%%%%%%%%%%%%%%%%%%%%%%%%%%%%

\subsection{Exact Scattering Amplitude}\label{subsec:exato}

Our goal is to derive the exact form of the first (non-renormalized) version of the scattering amplitude, by performing an explicit calculation within the phase-shifts method presented in \ref{sec:esp_2d}  and using  Eq. (\ref{1}), i.e.
\begin{eqnarray*}
\psi(r,\varphi)\approx e^{ikx}+f_k(\varphi)\frac{e^{ikr}}{\sqrt{-ir}}, \ x=r\cos\varphi,\  kr\gg1.\label{1_2}
\end{eqnarray*}
as a boundary condition. The normalization coefficient of the solutions of the cylindrical equation  (\ref{R}) and of the  phase-shifts are determined by the solution (\ref{Be}) in the limit $kr\gg1$ corresponding to  the specific choice\footnote{This ensures a regular solution at the origin. The generalization of this choice is the central core of section \ref{sec:ren}.} $B_m=0$,
\begin{eqnarray}
R_m(r)&\approx& A_m\sqrt{\frac{2}{kr}}\sin\left(kr+\frac{\pi}{4}-\frac{\pi}{2}\nu_m\right),\nonumber\\
&=&A_m\sqrt{\frac{2}{kr}}\sin\left(kr-\frac{\pi}{2}\left(m-\frac{1}{2}\right)+\left(m-\frac{\pi}{2}\nu_m\right)\right).
\end{eqnarray}
Comparing with  Eq. (\ref{2}) we see that  $Q_m(r)=J_{\nu_{m}}(kr)$, i.e. $A_m=1$, and 
\begin{eqnarray}
\delta_m=\frac{\pi}{2}\left(m-\nu_m\right).
\end{eqnarray}
It is worthwhile  to mention that the presence of the magnetic field breaks down the symmetry $\delta_m=\delta_{-m}$, that is usually  manifest in the case of central forces\footnote{Likewise, the degeneracy $m\rightarrow-m$ in the energy spectrum of a particle trapped in a ring is broken when a magnetic flux is passing inside the ring.}. Replacing the formula above in Eq. (\ref{3}) completely determines the exact  and single-valued form of the total wave function:
\begin{eqnarray}
\psi(r,\varphi)=\sum_{m=-\infty}^{\infty}(-1)^m(-i)^{\nu_{m}}J_{\nu_m}(kr).\label{sol_U}
\end{eqnarray}
This result turns out to be different from the original AB \emph{ansatz}\footnote{For a comprehensible derivation of the AB wave function  see ref. \cite{berry_eleg}.} by a factor of  $(-1)^m$ and it perfectly agrees with the wave function obtained in ref. \cite{Sakoda_AB} by two different methods. By construction, the wave function has the plane wave $e^{ikx}$ as a limit when $\nu_m\rightarrow|m|$, which is given by the equation (\ref{livre}). To make the calculations clearer we will introduce a new definition for the magnetic flux, $\Phi/\Phi_0\equiv\alpha=n+\zeta$, where $n\in\mathbb{N}$ and $0\le\zeta<1$. In this  notation we have
\begin{eqnarray}
\delta_m=\left\{ \begin{array}{ll} 
-\frac{\pi n}{2}-\frac{\pi}{2}\zeta, \ m+n+\zeta\ge0,\\
\pi m+\frac{\pi n}{2}+\frac{\pi}{2}\zeta, \ m+n+\zeta<0.
\end{array} \right.\label{delta_AB}
\end{eqnarray}
The phase-shifts  given by  Eq. (\ref{f_k}), together with the redefinition $m\rightarrow m-n$, gives
\begin{eqnarray}
f_k(\varphi)&=&e^{-in\varphi}\frac{1}{i\sqrt{2\pi k}}\sum_{m=-\infty}^{\infty}\left(e^{2i\delta_{m-n}}-1\right)e^{im\varphi},\nonumber\\
&=&e^{-in\varphi}\frac{1}{i\sqrt{2\pi k}}\left(\sum_{m=0}^{\infty}\left(e^{-in\pi}e^{-i\pi\zeta}-1\right)e^{im\varphi}+\sum_{m=1}^{\infty}\left(e^{-in\pi}e^{i\pi\zeta}-1\right)e^{-im\varphi}\right),\nonumber\\
&=&-e^{-in\varphi}\frac{2}{\sqrt{2\pi k}}\Bigg(e^{-i\frac{\pi}{2}(n+\zeta)}\sin\left(\frac{\pi}{2}(n+\zeta)\right)\sum_{m=0}^{\infty}e^{im\varphi}\nonumber\\
&&+e^{-i\frac{\pi}{2}(n-\zeta)}\sin\left(\frac{\pi}{2}(n-\zeta)\right)\sum_{m=1}^{\infty}e^{-im\varphi}\Bigg).\label{PG_f}
\end{eqnarray}
The above geometric progression series cannot be performed directly, however. First, one should  introduce an appropriate regularization procedure $e^{\pm im\varphi}\rightarrow e^{(\pm i\varphi-\varepsilon)m}$, and then take the limit $\varepsilon\rightarrow0$  at the end of calculations only, by making use of 
\begin{eqnarray}
\lim_{\varepsilon\rightarrow0}\frac{1}{x-a\pm i\varepsilon}=P\frac{1}{x-a}\mp i\pi\delta(x-a).
\end{eqnarray}
After some algebraic manipulations we get the final answer
\begin{eqnarray}
f_k(\varphi)\!\!\!&=&\!\!\!-\frac{i(-1)^ne^{-i(n+1/2)\varphi}}{\sqrt{2\pi k}}P\frac{\sin(\pi\zeta)}{\sin(\varphi/2)}-i\sqrt{\frac{2\pi}{k}}\left((-1)^n\cos\pi\zeta-1\right)\delta(\varphi),\nonumber\\
&=&-\frac{ie^{-i(n+1/2)\varphi}}{\sqrt{2\pi k}}P\frac{\sin\left(\pi\frac{\Phi}{\Phi_0}\right)}{\sin(\varphi/2)}-i\sqrt{\frac{2\pi}{k}}\left[\cos\left(\pi\frac{\Phi}{\Phi_0}\right)-1\right]\delta(\varphi),\nonumber\\
&=&\sqrt{\frac{2}{\pi k}}\left\{\sin\left(\pi\frac{\Phi}{\Phi_0}\right)P\frac{z^{-n}}{z-1}-i\left[\cos\left(\pi\frac{\Phi}{\Phi_0}\right)-1\right]\delta(z-1)\right\},\label{f_U}
\end{eqnarray}
where $z=e^{i\varphi}$. For $z\neq1$, i.e. without taking the Dirac delta function term into account, the above formula  differs from the original AB result by a phase factor, while it coincides up to constant phases with the ones obtained in refs. \cite{espalhamento_AB_WKB,cilindro_AB1,cilindro_AB2} in the limit of a vanishing cylinder radius. It is important to note that  all the calculations presented in refs. \cite{espalhamento_AB_WKB,cilindro_AB1} were made in the WKB approximation, thus making it  evident   the semiclassical nature of the considered scattering process. Finally, for $z\neq1$ we obtain the so called AB cross-section
\begin{eqnarray}
\sigma_k(\varphi)=|f_k(\varphi)|^2=\frac{1}{2\pi k}\frac{\sin^2\left(\pi\frac{\Phi}{\Phi_0}\right)}{\sin^2(\varphi/2)},\label{s_AB1}
\end{eqnarray}
whose most remarkable property is its periodic dependence on the magnetic flux, which leads to a null cross-section when $\Phi/\Phi_0\in\mathbb{N}$.

Now we turn our attention to the $\delta$-term, which is absent in the original AB paper, but it  is not a novelty indeed  \cite{Ruji_AB}. It is clearly related to the non-convergence of the geometric  series in (\ref{PG_f}) for $z=1$, whose net effect is the divergence of the total cross-section, in agreement with the optical theorem\footnote{Which also agrees with the divergence of the integration of (\ref{s_AB1}) on the two-dimensional solid angle.} $\sigma_t\sim\Im m(f(0))$.  It represents a non-trivial quantum effect, since the divergences of the total cross-section are usually seen as a signature for a long-range interaction, e.g. the Coulomb-like scattering. The $\delta$-term can be understood, at least partially, as a  ``side effect'' behind the displacement of the phase $e^{i\alpha\varphi}$ from the incident wave to the scattered one as pointed out in ref.  \cite{Hagen_AB}
\begin{eqnarray}
2ie^{\frac{\alpha}{2}(\pi-\varphi)}\sin\left(\frac{\alpha}{2}(\pi-\varphi)\right)e^{ikr\cos\varphi}\stackrel{r\rightarrow\infty}{\longrightarrow}\sqrt{\frac{2\pi}{ikr}}(\cos\alpha-1)e^{ikr}\delta(\varphi).\label{limit}
\end{eqnarray}
But one can instead advocate that such an explanation is not sufficient, and try to interpret this term as an error within the phase-shift method, associated to the improper interchange of the limit $r\rightarrow\infty$ with the summation over $m$ \cite{Hagen_AB}. The problem arises because $\delta_m$ is $m$-independent, while it was expected that it goes to zero for large $m$ in order to ensure the convergence of the series. On the other hand, using a couple of different exact approaches, the same equation (\ref{s_AB1}) appears in ref. \cite{Sakoda_AB} and,  if we take into account the unitarity of the S-matrix, the $\delta$-term cannot be neglected, as it was pointed out by Sakoda \emph{et al.} \cite{Sakoda_AB}. Therefore, one might  consider  the phase-shifts method to be equally appropriate for treating the AB scattering, since all the other approaches (including the original AB one) are known to suffer from the very same problem on the $\varphi=0$ line. It, however, has an advantage of not making use of complicated asymptotic relations involving Bessel functions, and to lead to a single valued wave function, since the non-integrable phase factor only appears asymptotically at the limit (\ref{limit}).

\vspace{0,5cm}

Let us summarize: the exact cross-section of the AB infinitely thin solenoid problem was derived  by using the b.c's  specific for  the short-range two-dimensional elastic scattering, i.e. within the frameworks of the phase-shifts method. Although we did not use the standard  minimal coupling form of the incident wave function (\ref{free_AB}), our formula (\ref{sol_U})  by construction represents a single-valued solution of the Schr\"odinger equation and it also reproduces (\ref{1}) in the limit $r\rightarrow\infty$.

\section{AB Scattering renormalization}\label{sec:ren}
In this section we will analyze more carefully all the consequences  of the b.c. $B_m=0$, when imposed on the solutions of equation (\ref{Be}). Our first step is to rewrite Eq. (\ref{R}) in terms of a new function $u_m(r)=\sqrt{r}R_m(r)$
\begin{eqnarray}
&&-u''_m(r)+\frac{\mathcal C_m}{r^2}u_m(r)=k^2u_m(r), \ \mathcal C_m=\nu_m^2-1/4\ge-1/4,\label{Cal}\\
&&u_m(r)=\sqrt{r}R_m(r)=\sqrt{r}A_mJ_{\nu_m}(kr)+B_m\sqrt{r}N_{\nu_m}(kr),\label{U}
\end{eqnarray}
which turns out to coincide with the wave function of a particle interacting with the one-dimensional Calogero potential \cite{Cal1,Cal3}, which is also called \emph{conformal potential}, $V_m(r)=\hbar^2\mathcal C_m/(2Mr^2)$. The formal equivalence between the Calogero problem and the AB scattering is indeed well known in the literature \cite{Gitman} and it will serve for us as a guide for the investigations  presented in this section.

The quantization of singular potentials is not a straightforward process since  the standard   quantum mechanical treatment might give rise to  non self-adjoint (s.a.) Hamiltonian operators. This is exactly what happens in the case of the Calogero potential for $\mathcal C_{m}<3/4$. Detailed studies of the s.a. extensions of the Calogero/AB Hamiltonian operators can be found in refs.  \cite{Gitman,Gitman2,Cal_s.a.}. Instead of applying  the s.a. extensions techniques however, we will face the problem by using another equivalent approach known as the  Renormalization method of Beane at al. \cite{Beane}. By implementing it within  the AB scattering context, we will be able to  derive the renormalized phase-shifts and the exact form of the corrections to the AB cross-section (\ref{s_AB1}), including its non-zero value when the magnetic flux is quantized. 

\subsection{Renormalization as self-adjoint extension}
%------------------------------------------
\begin{figure}[ht] 
\centering
\subfigure[]{
\includegraphics[scale=0.24]{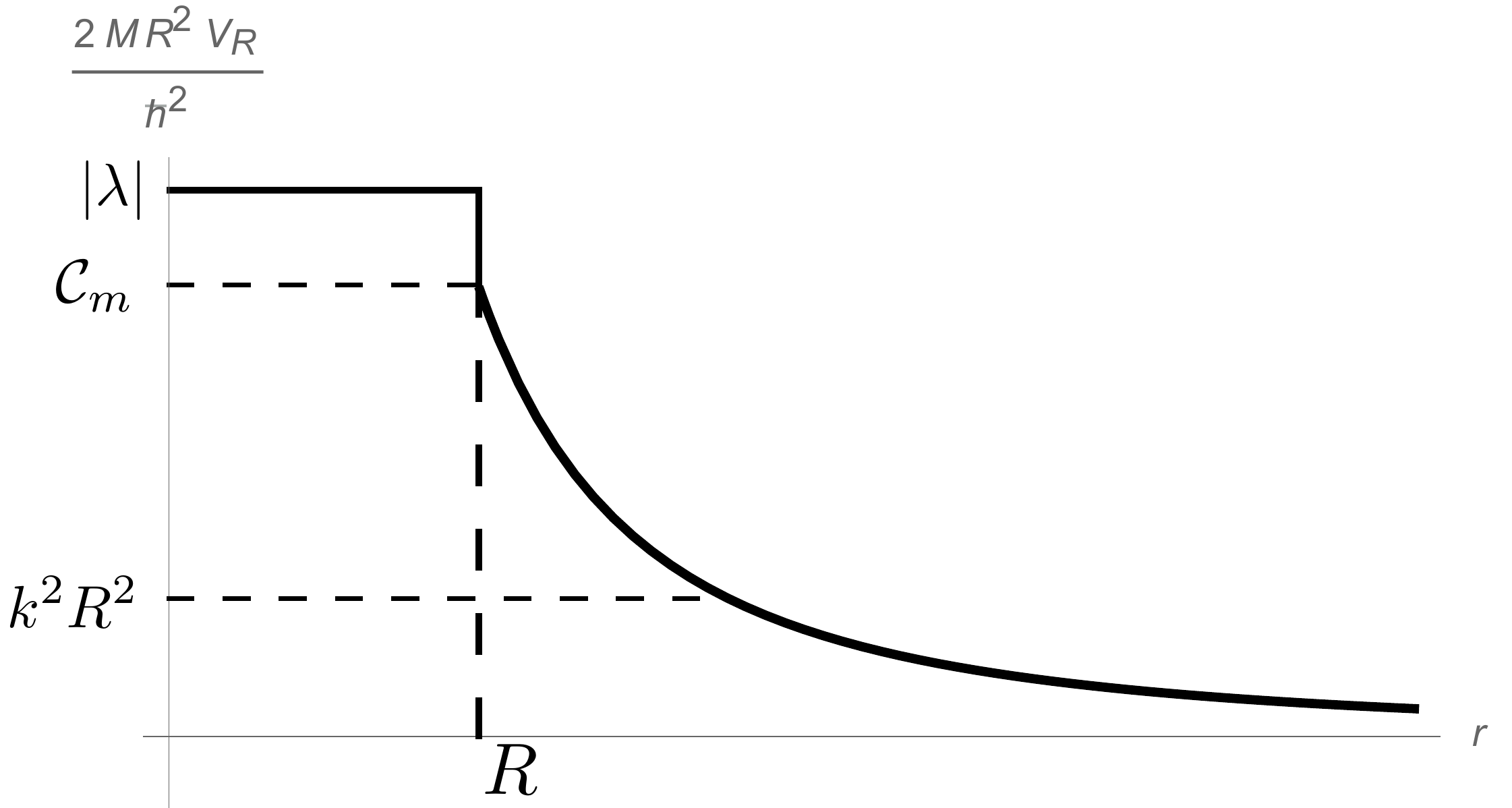}
\label{fig:VR_rep_rep}
}
\subfigure[]{
\includegraphics[scale=0.24]{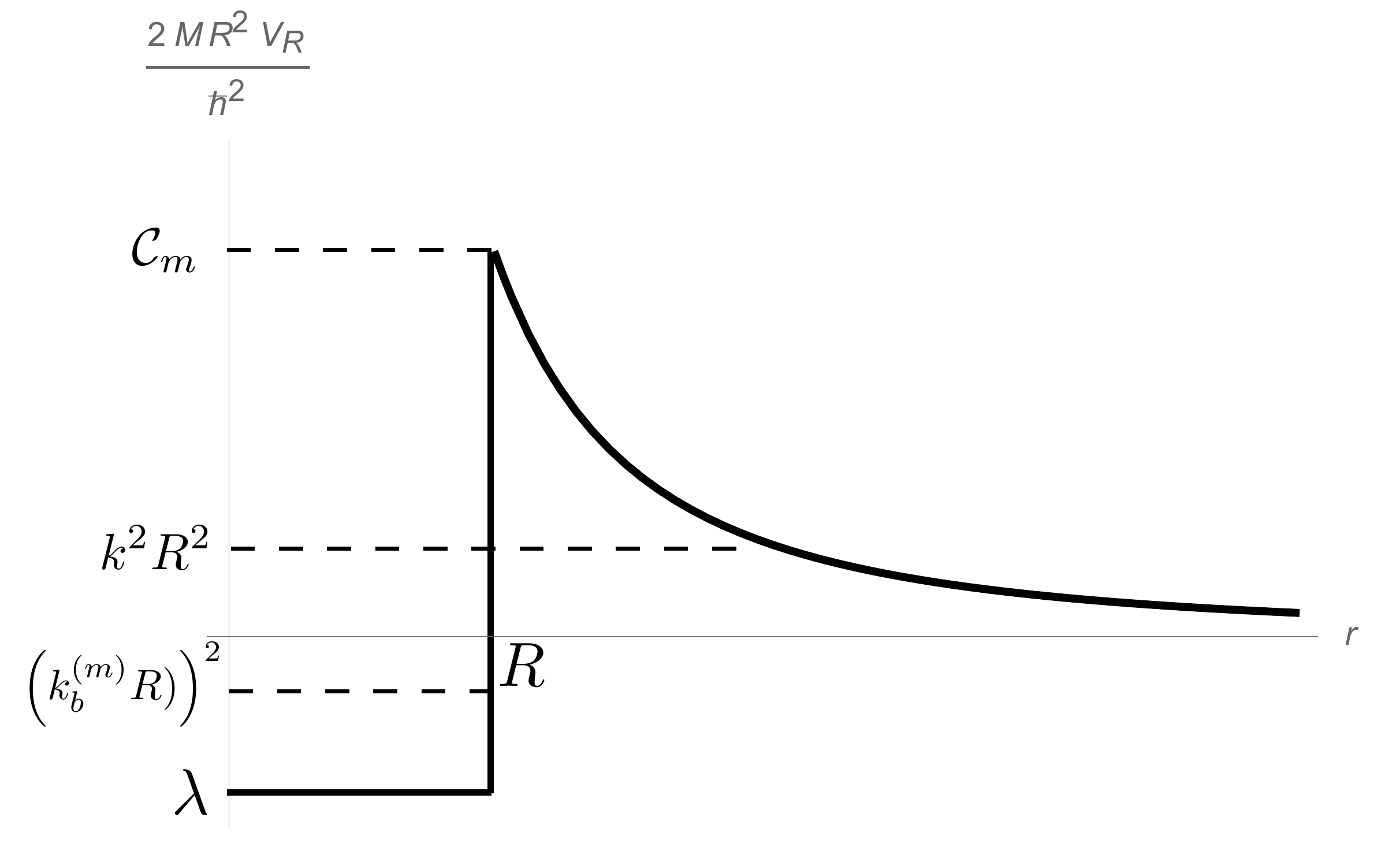}
\label{fig:VR_rep_at}
}
\caption{Regularized potential in repulsive case. In the region $r<R$ we have: (a) barrier potential, $\lambda<0$; (b)  finite well potential, $\lambda>0$.}
\label{fig:VR_rep}
\end{figure}
%------------------------------------------
The justification for choosing $B_m=0$ in Eq. (\ref{Be}) comes from the hypothesis that the cylindrical wave functions must be regular at the origin of the coordinate system.  This is a sufficient but not necessary condition; in fact, we only need the function $R_{m}(r)$ to be square-integrable at the origin, i.e. $\int^rdr'r'^2R_m(r')=\int^rdr'u_m(r')<\infty$ when $r\rightarrow0$. For $\mathcal{C}_m\ge 3/4$ the result is exactly the same as assumed in the previous section ($B_m=0$), while within  the range $0\le\nu_m<1$, i.e. $-1/4\le \mathcal C_m<3/4$,  both  solutions in  (\ref{Be}) are in fact square-integrable. The region $\mathcal C_m<-1/4$ is not directly related to the AB scattering, it deserves however  a separate consideration due to the many important physical implications, like the Efimov effect\footnote{A modern review on the subject can be found in \cite{Efimov_rev}.} \cite{Efimov,Efimov2}, and other related  applications \cite{Ren_1,Arg1,Ren_2,Ren_3,Ren_4,Ren_5}.

Our investigation will be focused on the range of  ``medium-weak'' values of the coupling  $-1/4\le \mathcal C_m<3/4$ ($0\le\nu_m<1$), where a precise (renormalized) definition of the AB Hamiltonian is needed. This interval appears in the AB scattering problem in two different cases: i) $m=-n$, then $\nu_{-n}=\zeta$; ii) $m=-n-1$, with $\nu_{-n-1}=1-\zeta$.
Here $\zeta\neq0$. Notice that in the $\zeta=1/2$ case a  degeneracy $\nu_{-n}=\nu_{-n-1}=1/2$ occurs, which corresponds to the null Calogero potential, i.e.  $\mathcal{C}_{-n}=\mathcal{C}_{-n-1}=0$. When the magnetic flux is quantized, $\zeta=0$, the ``medium-weak''-range contains only  the mode $m=-n$, with $\nu_{-n}=0$ ($\mathcal C_{-n}=-1/4$). It turns out that for all these modes the requirement of  square-integrability at the origin does not select the AB solution in an unique way (neglecting a multiplicative constant factor). It leaves  the constants $A_m$ and $B_m$  arbitrary and thus introduces a new length scale in the AB problem. The correct choice of the b.c.'s at $r=0$ for all these cases can be achieved  by applying the methods of ref. \cite{medium_AB}. In order to make this point as clear as possible, we  first consider the simplest case of  zero energy solutions of Eq. (\ref{Cal}):
\begin{eqnarray}
u_m^{(0)}(r)=c_+^{(m)}r^{1/2+\nu}+c_-^{(m)}r^{1/2-\nu}=c_+^{(m)}r^{1/2+\nu}\left(1+\varepsilon\left(\frac{L_m}{r}\right)^{2\nu}\right),\label{U0}
\end{eqnarray}
where $c_-^{(m)}/c_+^{(m)}\equiv \varepsilon L_m^{2\nu}$. We have denoted by $\varepsilon=\pm1$ the sign of the ratio, while  $L_m\ge0$ is the proper length scale parameter\footnote{All quantities have the subindex ``$m$'', since each $m$ mode corresponds to a different Calogero model.}. Classically, the Calogero potential is known to be  a conformally invariant model, i.e. without any scales involved. The quantization process   introduces,  for certain medium-weak couplings, a new $L_m$ scale even for the zero energy solutions. Although it  breaks the rescaling symmetry, one should mention that  the conformal symmetry is in fact preserved  when one of the solutions, $c_-^{(m)}=0$ ($L_m=0$) or $c_+^{(m)}=0$ ($L_m\rightarrow\infty$) is discarded.  All the physical observables are now functions of the ratio $c_-^{(m)}/c_+^{(m)}$, thus creating ``quantum families'' of one (for $m=-n$, $\zeta=0$) or two (for $m=-n$ and $m=-n-1$ with $\zeta\neq0$) parameters, whose precise values can be eventually determined by measurement.

\begin{figure}[ht] 
\centering
\subfigure[]{
\includegraphics[scale=0.25]{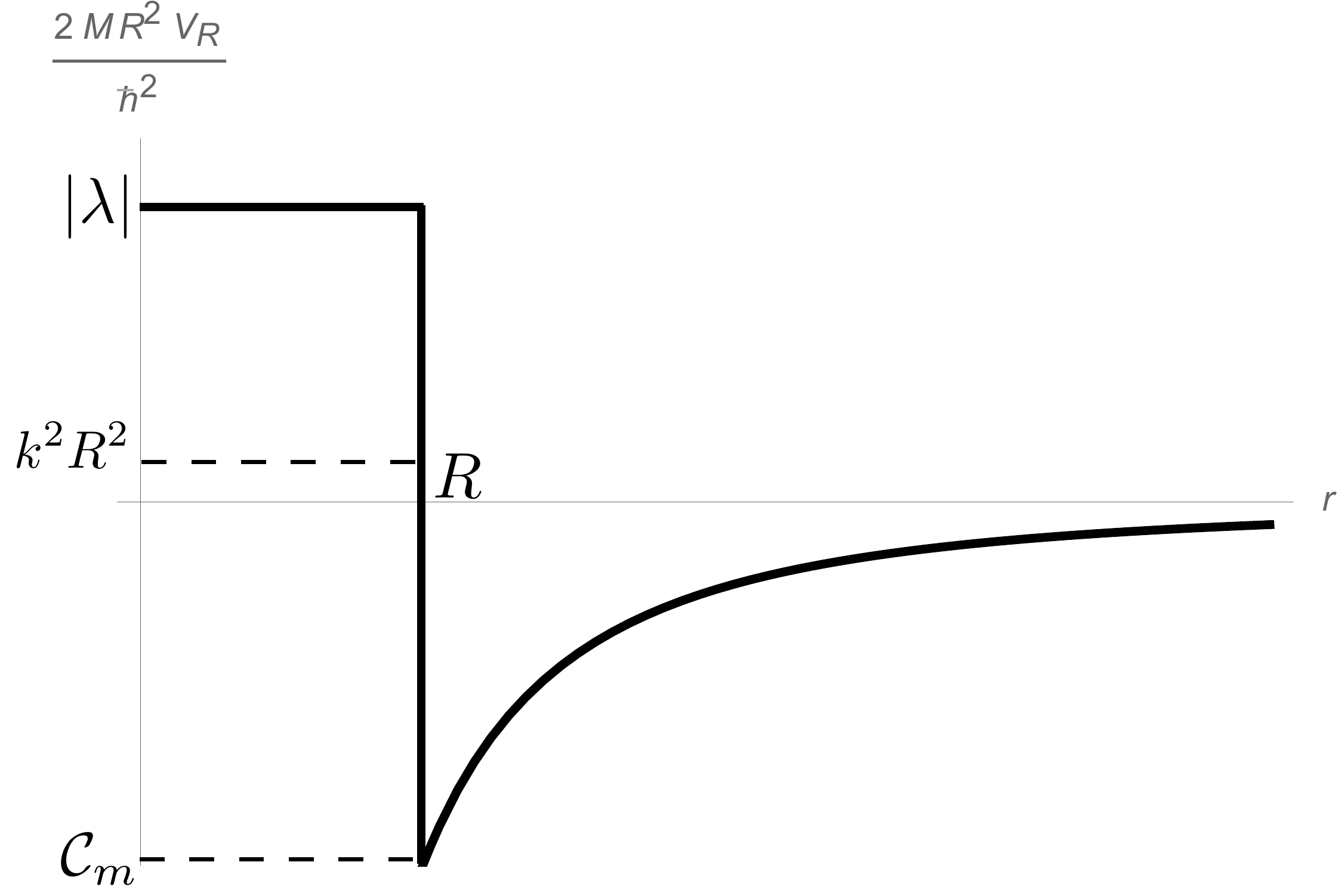}
\label{fig:VR_at_rep}
}
\subfigure[]{
\includegraphics[scale=0.25]{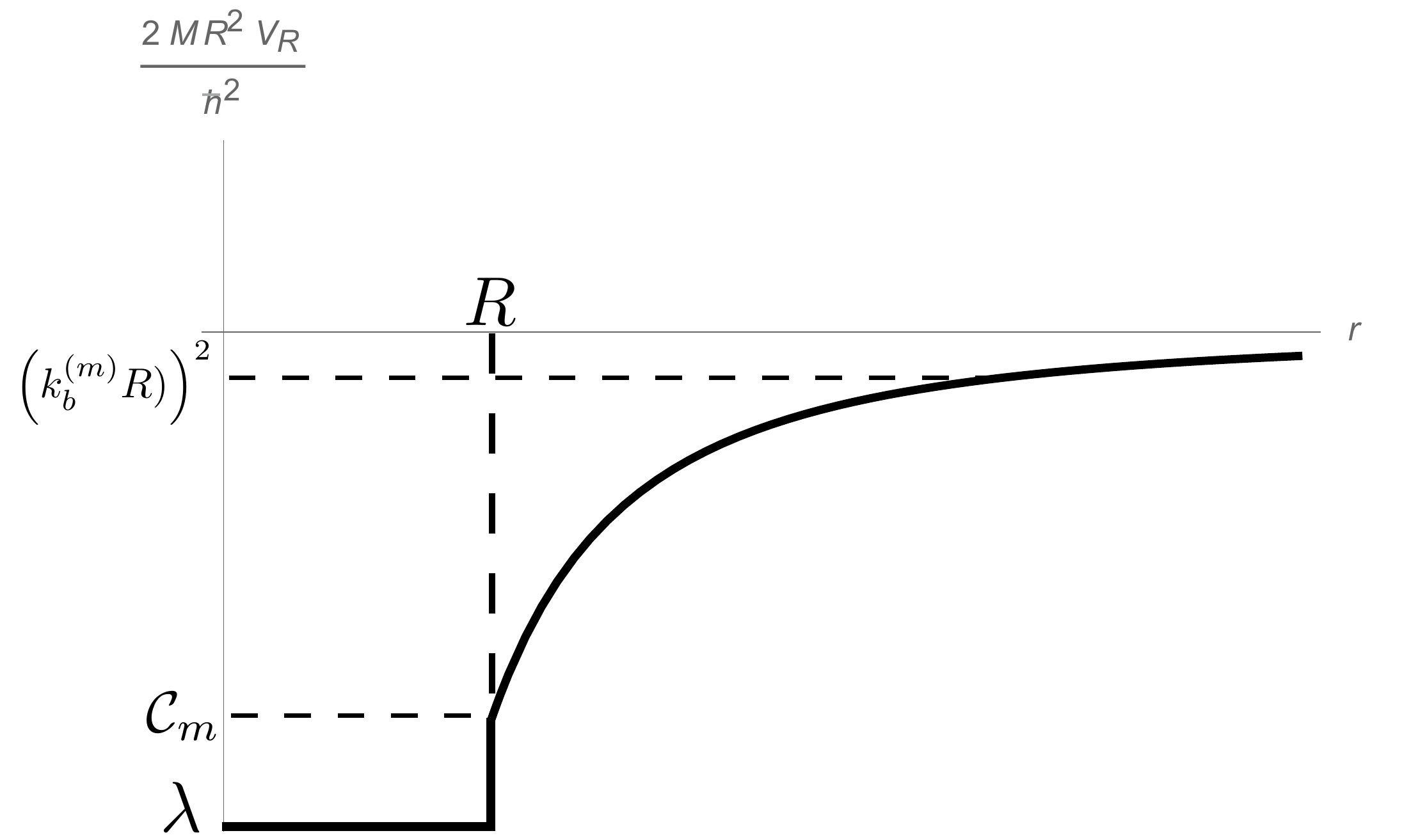}
\label{fig:VR_at_at}
}
\caption{Regularized potential in attractive case. In the region $r<R$ we have: (a) A barrier potential, $\lambda<0$; (b) A well potential, $\lambda>0$.}
\label{fig:VR_at}
\end{figure}

Once the $\varepsilon$ and $L_m$ parameters are given, in order to impose the correct Dirichlet\footnote{Other choices can be also made, such as the Robin (or mixed) boundary condition used in \cite{espalhamento_AB_WKB}.} b.c. that relate  $A_m$ e $B_m$  in a consistent way,  we have to  regularize the potential by introducing a ``short distance cutoff''\footnote{The ``short distance cutoff' term precisely means $kR\ll1$ for all the non zero energy solutions.}, say at $r=R$. In the interior region, the potential shape is modified in order to make it regular at the origin, and only then the condition $u_m(0)=0$ is imposed. Usually, the potential for $r<R$  is replaced by a simpler one, such as a well or a barrier,  or (equivalently)  a Dirac delta function. The final physical predictions turn out to be independent of the regularization method adopted, thus ensuring the universality of the final results. In this paper our option is to use the potential well/barrier, and thus the  regularized Calogero potential takes the form:
\begin{eqnarray}
\frac{2M}{\hbar^2}V_R(r)=\left\{ \begin{array}{ll} 
 \frac{\mathcal{C}_m}{r^2}, \ r>R,\\
-\frac{\lambda}{R^2}, 0<r<R.
\end{array} \right. 
\end{eqnarray}
Generally the coupling $\lambda$ is considered positive (a well potential), see for example \cite{medium_AB}. There is no need, however, to impose  this restriction, and we will further  assume that $\lambda$ is a  free parameter --- for  negative values, it corresponds to a barrier potential. Figures \ref{fig:VR_rep} and \ref{fig:VR_at} represent all the possibilities for the regularized potential. The first one describes the two curves for the repulsive interaction, $0\le\mathcal{C}_m<3/4$ ($1/2\le\nu_m<1$), while the second one shows the two curves in the attractive case  $-1/4\le\mathcal{C}_m<0$ ($0\le\nu_m<1/2$).

\begin{figure}[ht]
    \centering
    \includegraphics[scale=0.32]{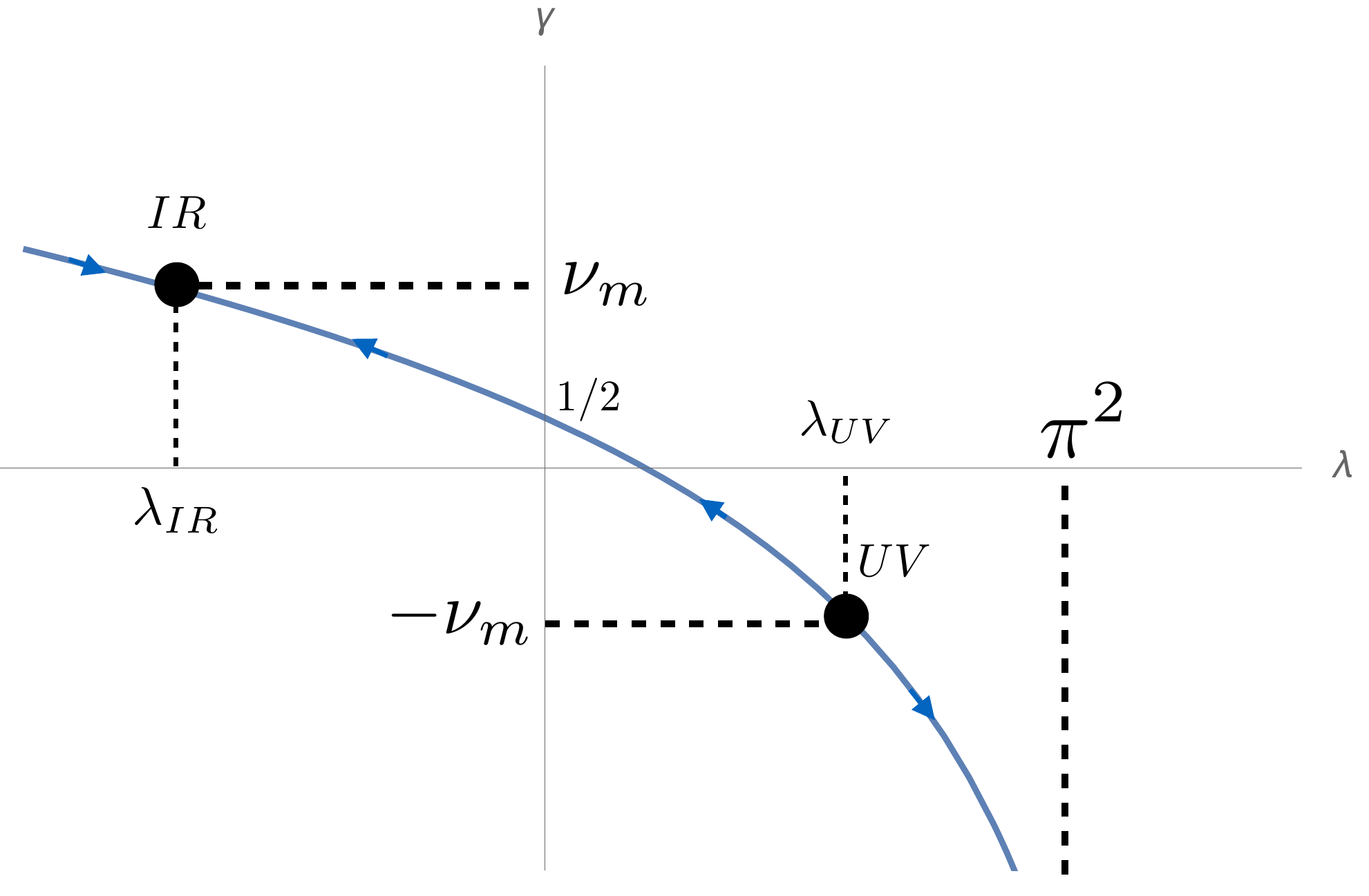}
    \begin{quotation}
    \caption[ed]{\small The ``running coupling'' $\gamma$ as a function of the strength coupling $\lambda$. Within the range $-\infty<\lambda<\sqrt{\pi}$ the parameter $\gamma$ assumes all the values on the line. The arrows indicate the directions of the RG flows, that will be discussed in section \ref{sec:beta}. The solid dots are the IR and UV fixed points.
}
     \label{fig:gamma_la}
     \end{quotation}
\end{figure}

With the above choice for the regularized potential, one can easily derive the most general  $E>0$ solutions, which  satisfy the  $u_m(0)=0$ condition:
\begin{eqnarray}
u_m(r)=\left\{ \begin{array}{ll} 
\sqrt{r}A_mJ_{\nu_m}(kr)+B_m\sqrt{r}N_{\nu_m}(kr), \ r>R,\\ 
\left[\sqrt{R}A_mJ_{\nu_m}(kR)+\sqrt{R}B_mN_{\nu_m}(kR)\right]\frac{\sin \rho r}{\sin\rho R}, \ \rho=\frac{1}{R}\sqrt{\lambda+(kR)^2},\ \ r<R.
\end{array} \right.
\end{eqnarray}
%------------------------------------------
%------------------------------------------
For the specific case with $E=0$, the requirement for continuity  of the wave function and its first derivative leads to the following equality
%No caso de $E=0$, as continuidades da fun\c c\~ao de onda e de sua primeira derivada levam a seguinte igualdade:
\begin{eqnarray}
\gamma(R)=\nu_m\frac{(1-\varepsilon(L_m/R)^{2\nu_m})}{(1+\varepsilon(L_m/R)^{2\nu_m})},\label{gamma_R}
\end{eqnarray}
with
%com
\begin{eqnarray}
\gamma(\lambda)\equiv \left\{ \begin{array}{ll} 
 \sqrt{\lambda}\cot\sqrt{\lambda}-\frac{1}{2}, \ \lambda>0,\\
\sqrt{|\lambda|}\coth\sqrt{|\lambda|}-\frac{1}{2}, \lambda<0,
\end{array} \right. \label{gamma_la}
\end{eqnarray}
where $\gamma$ is named as  the ``running coupling constant".  We shall restrict our analysis to the range\footnote{For $\lambda>\pi^2$, the function $\gamma(\lambda)$ exhibits a periodic behavior, describing the strong attractive case, $C_m<-1/4$, which is not of our interest here.} $-\infty<\lambda<\pi^2$ only, where $\gamma(\lambda)$ is a continuous ($\gamma(0^{\pm})=1/2$) and single-valued function defined on the entire  real line, as shown in  fig.  \ref{fig:gamma_la}. These requirements assure that  the equalities (\ref{gamma_R}) and (\ref{gamma_la}) determine the coupling $\lambda$ as a function of the cutoff $R$.  

The explicit knowledge of $\gamma(\lambda)$  and $\lambda(R)$ allows us to directly apply the standard field-theoretical and statistical mechanics  renormalization group  approach to the case of the AB scattering. As usual, the most important quantity is given by the derivative $d\gamma/d(\ln(R/L_m))$, representing   the beta function, $\beta(\gamma)$, of the regularized AB model and the following  section \ref{sec:beta}  is devoted to its study.  As one can easily verify in the  two limits of $L_m$ the coupling $\gamma$ becomes an $R$-independent constant. The first one is $L_m=0$ ($c_-=0$), $\gamma\equiv\gamma_{IR}=\nu_m$ and  $L_m\rightarrow\infty$ ($c_+=0$), $\gamma\equiv\gamma_{UV}=-\nu_m$ is the other one. These are, respectively, the IR and UV fixed points of the model, where only one solution holds and the conformal symmetry is restored. It is assumed  that  $\lambda(R)$ remains unchanged under  the renormalization procedure, i.e. it  is  given by Eq. (\ref{gamma_R}) for all  the energies. 

We next consider the $E>0$ case\footnote{The possibility of solutions for $E<0$, i.e. bound states, will be considered in subsection \ref{subsec:ligado}.}. As the continuity of the logarithmic derivative $u'_m(r)/u_m(r)$ is required at $r=R$, with $kL\ll1$, we only need the asymptotic expressions
%%%%%%%%%%%%%%%%%%%%%%%FIGURA%%%%%%%%%%%%%%%
%%%%%%%%%%%%%%%%%%%%%%%%%%%%%%%%%%%%
\begin{eqnarray*}
J_{\nu}(z)\approx\frac{1}{\Gamma(1+\nu)}\left(\frac{z}{2}\right)^{\nu}\, \ \ N_{\nu}(z)\approx-\frac{\Gamma(\nu)}{\pi}\left(\frac{z}{2}\right)^{-\nu}, \ z\ll1,
\end{eqnarray*}
leading to
\begin{eqnarray*}
\rho R\cot(\rho R)-1/2\approx  \gamma=\nu_m\frac{\frac{A_m}{\Gamma(1+\nu_m)}\left(\frac{kR}{2}\right)^{\nu_m}+B_m\frac{\Gamma(\nu_m)}{\pi}\left(\frac{kR}{2}\right)^{-\nu_m}}{\frac{A_m}{\Gamma(1+\nu_m)}\left(\frac{kR}{2}\right)^{\nu_m}-B_m\frac{\Gamma(\nu_{m})}{\pi}\left(\frac{kR}{2}\right)^{-\nu_m}}.
\end{eqnarray*}
By eliminating $\gamma$ with  the help of  Eq. (\ref{gamma_R}), one can easily derive the explicit form of the ratio $B_m/A_m$:
\begin{eqnarray}
\frac{B_m}{A_m}=\frac{(-\varepsilon)\pi}{\Gamma(\nu_m)\Gamma(1+\nu_m)}\left(\frac{kL_m}{2}\right)^{2\nu_{m}}, \ 0<\nu_m<1.\label{B/A}
\end{eqnarray}
Once the renormalization process has been completed, the ratio of the constants $A_m$ and $B_m$ have been uniquely fixed  as a  function of the length scale $L_m$ only, i.e. independent of the ``cut-off'' scale $R$, as expected. Then the  renormalized cylindrical wave function for the modes $m=-n$ or $m=-n-1$  takes the form:
\begin{eqnarray}
R^{ren}_m(r)=A_mJ_{\nu_m}(kr)+B_mN_{\nu_m}(kr), \ r>R,
\end{eqnarray}
where the constants $A_m$ and $B_m$  are now related by Eq. (\ref{B/A}). In the limit $kr\gg1$ we get
\begin{eqnarray}
R^{ren}_m(r)\approx \sqrt{\frac{2}{\pi kr}}\left[A_m\cos\left(kr-\frac{\pi}{4}-\frac{\pi}{2}\nu_m\right)+B_m\sin\left(kr-\frac{\pi}{4}-\frac{\pi}{2}\nu_m\right)\right].
\end{eqnarray}
 By comparing with the boundary condition (\ref{2}), required by the elastic scattering, we realize that
$$R^{ren}_m(r)\approx \sqrt{\frac{2}{\pi kr}}\sin\left(kr-\frac{\pi}{2}\left(m-\frac{1}{2}\right)+\frac{\pi}{2}\left(m-\nu_m\right)+\Sigma_m\right),$$
where we have introduced  the parameterization $ A_m=\cos\Sigma_m$ and $B_m=-\sin\Sigma_m$ in terms of the correction $\Sigma_m $ to the  phase shifts, 
\begin{eqnarray}
\tan\Sigma_m(k)=\frac{\varepsilon\pi}{\Gamma(\nu_m)\Gamma(1+\nu_m)}\left(\frac{kL_m}{2}\right)^{2\nu_{m}}.
\end{eqnarray}
 The above results  demonstrate that the renormalization of the Calogero potential (for fixed $m$) generates a one-parameter  family of models, i.e. exactly the same conclusion as the one  obtained in \cite{Gitman} by  the self-adjoint extensions methods. Finally, we find as a consequence of the above equations the explicit form of the new renormalized phase-shifts ($\zeta\neq0$):
\begin{eqnarray}
\delta^{ren}_m=\left\{ \begin{array}{ll} 
 \frac{\pi}{2}\left(m-\nu_m\right)+\Sigma_m , \ m=-n, \ m=-n-1,\\
\frac{\pi}{2}\left(m-\nu_m\right), \textrm{others values of $m$}.
\end{array} \right.
\end{eqnarray}
Let us remind that if $\zeta = 1/2 $, $\nu_{n} = \nu_{n-1} = 1/2 $, and the two scales do coincide, $L_n = L_{-n-1}$, then a degeneracy for the phase-shifts corrections $\Sigma_n = \Sigma_{n-1}$ arises.

When the magnetic flux is quantized ($\Phi/\Phi_0\in\mathbb{N}$), i.e. for $\zeta=0$, the unique mode $m=-n$ needs to be renormalized. In this case  the renormalized solutions turn out to be  completely different, and all the process has to be reformulated. We start again with the zero energy case: 
\begin{eqnarray}
u_{-n}(r)= \left\{ \begin{array}{ll} 
 \sqrt{r}\left(1+\bar c\ln(r/L)\right), \ r>R,\\
\sqrt{R}\left(1+\bar c\ln(R/L)\right)\frac{\sin\left(\sqrt{\lambda}r/R\right)}{\sin\sqrt{\lambda}}, \ r<R,
\end{array} \right. 
\end{eqnarray}
where $\bar c$ is a dimensionless constant and $L$ defines again the length scale. The continuity of $u'_m(r)/u_m(r)$ now implies:
\begin{eqnarray}
\gamma(R)=\frac{\bar c}{1+\bar c\ln(R/L)},\label{gamma_0}
\end{eqnarray}
with the same $\gamma(\lambda)$ as in Eq. (\ref{gamma_la}). We next reconsider  the renormalization procedure for $E>0$:
\begin{eqnarray}
&u_{-n}(r)=\left\{ \begin{array}{ll} 
 A_{-n}\sqrt{r}J_{0}(kr)+B_{-n}N_{0}(kr), \ r>R,\\
\left[A_{-n}\sqrt{R}J_{0}(kR)+B_{-n}\sqrt{R}N_{0}(kR)\right]\frac{\sin(\rho r)}{\sin(\rho R)},\  \rho=\frac{1}{R}\sqrt{\lambda+(kR)^2} \ r<R.
\end{array} \right.
\end{eqnarray}
Again, the  continuity of $u_{-n}'(r)/u_{-n}(r)$, together with the hypothesis $kR\ll1$, yield 
\begin{eqnarray}
\gamma(\lambda)=\frac{\Theta}{1+\Theta\ln(kR)}, \ \ \Theta\equiv\frac{2}{\pi}\left(\frac{A_{-n}}{B_{-n}}+\frac{2C_E}{\pi}\right)^{-1},
\end{eqnarray}
where $C_E\approx0.5772$ is the Euler's constant and the  following asymptotic forms
$$J_0(z)\approx1+\mathcal{O}(z^2), \ \ \ N_0(z)\approx \frac{2}{\pi}\left(\ln(z/2)+C_E\right)+\mathcal{O}(\ln(z/2)z^2),$$
have been used. Finally, by eliminating $\gamma(\lambda)$ from the above expression 
\begin{eqnarray}
\frac{A_{-n}}{B_{-n}}=\frac{2}{\pi}\left(\frac{1}{\bar c}-C_E-\ln(kL)\right)=-\frac{2}{\pi}\ln(k\bar{L}), \ \bar{L}=e^{-\frac{1}{\bar c}+C_E}L,\label{A_0}
\end{eqnarray}
the real physical prediction (without any cutoff $R$-dependence)  is obtained. The corresponding  renormalized cylindrical wave function now takes the asymptotic form 
\begin{eqnarray}
R^{ren}_{-n}(r)\approx \sqrt{\frac{2}{\pi kr}}\left(A_{-n}\cos(kr-\pi/4)+B_{-n}\sin(kr-\pi/4)\right), \ kr\gg1.
\end{eqnarray}
With an appropriate definition
\begin{eqnarray}
A_{-n}=\cos\bar\Sigma, \ \ B_{-n}=-\sin\bar\Sigma,
\end{eqnarray}
we have again a one-parametric family of consistent models. Observe that the Eq. (\ref{2}) is obtained once more,  but with  the new form of the phase-shifts
\begin{eqnarray}
\delta^{ren}_{-n}=-\frac{n\pi}{2}+\bar{\Sigma}(k),\ \ \cot(\bar{\Sigma}(k))=\frac{2}{\pi}\ln(k\bar{L}).
\end{eqnarray}

The calculations of the renormalized phase-shifts for all the cases of interest (i.e. for all allowed values of $\zeta$), that by construction are cut-off independent, provides a firm base for the  construction of the physically relevant and observable quantities, which in the context of the AB problem  are in fact reduced to the knowledge of the corrections to the original AB elastic scattering  cross-section. 

\subsection{Renormalized scattering amplitudes}

Once we have the renormalized phase-shifts, all the scattering data is at our hands. What remains is  to derive   the renormalized scattering amplitudes. Let us start with the case $\zeta\neq0$:
\begin{eqnarray}
f_k^{ren}=&&\frac{e^{-in\varphi}}{i\sqrt{2\pi k}}\sum_{m=-\infty}^{\infty}\left(e^{2i\delta_{m}^{ren}}-1\right)e^{im\varphi},\\
=&&f_k+(-1)^ne^{-in\varphi}\sqrt{\frac{2}{\pi k}}\left[e^{i\left(\Sigma_{-n}-\pi\zeta\right)}\sin\Sigma_{-n}+e^{i\left(\Sigma_{-n-1}+\pi\zeta-\varphi\right)}\sin\Sigma_{-n-1}\right],\nonumber\\
=&&\frac{(-1)^ne^{-i(n+1/2)\varphi}}{\sqrt{2\pi k}}\Bigg\{\frac{\sin(\pi\zeta)}{\sin(\varphi/2)}+\cos\left(\frac{\varphi}{2}-\pi\zeta+2\Sigma_{-n}\right)+\cos\left(\frac{\varphi}{2}-\pi\zeta-2\Sigma_{-n-1}\right)\nonumber\\
&&-2\cos\left(\frac{\varphi}{2}-\pi\zeta\right)+i\left[\sin\left(\frac{\varphi}{2}-\pi\zeta+2\Sigma_{-n}\right)-\sin\left(\frac{\varphi}{2}-\pi\zeta-2\Sigma_{-n-1}\right)\right]\Bigg\}
\end{eqnarray}
with $f_k$ given by Eq. (\ref{f_U}). The new cross-section $\sigma^{ren}_k=|f_k^{ren}|^2$ may be obtained by a direct calculation. Here we will present a simple version of it, valid at the low energy limit, where $\Sigma_{-n}(k),\Sigma_{-n-1}(k)\ll1$. For $\varphi\neq0$ we have
\begin{eqnarray}
\sigma_k^{(ren)}\approx \frac{1}{2\pi k}\frac{\sin^2\pi\zeta}{\sin^2(\varphi/2)}+\frac{2}{\pi }\frac{\sin\pi\zeta\sin(\varphi/2-\pi\zeta)}{\sin(\varphi/2)}\left(\frac{\Sigma_{-n-1}(k)-\Sigma_{-n}(k)}{k}\right).
\end{eqnarray}
The two terms of the first correction  turn out  to have a simple power-law behavior  $\Sigma_{-n}(k)/k\sim k^{2\zeta-1}$ and $\Sigma_{-n-1}(k)/k\sim k^{1-2\zeta}$. Therefore at  low enough energies, and  for $\zeta\neq1/2$, one contribution is always dominating over the other one, while  for $\zeta=1/2$ the above first correction is in fact energy independent.

For the configuration with $\zeta=1/2$ and $L_{-n}=L_{-n-1}\equiv L$, i.e. for $\Sigma_{-n}(k)=\Sigma_{-n-1}(k)\equiv\Sigma^{1/2}(k)=\arctan(kL)$, the exact result takes a rather simple and compact form ($\varphi\neq0$):
\begin{eqnarray}
\sigma^{ren}_k(\varphi)=\frac{1}{2\pi k\,\sin^2(\varphi/2)}\left[\left(1-4\sin^2\left(\Sigma^{1/2}(k)\right)\sin^2\left(\frac{\varphi}{2}\right)\right)^2+4\sin^2\left(2\Sigma^{1/2}(k)\right)\sin^4\left(\frac{\varphi}{2}\right)\right].\nonumber
\end{eqnarray}

The case $\zeta=0$ ($\Phi/\Phi_0\in\mathbb{N}$) is very special. The usual AB cross-section vanishes for $\varphi\neq0$, and all the contribution comes from the renormalization. After a simple calculation we get
\begin{eqnarray}
\bar\sigma_k^{ren}=\frac{2}{\pi k}\sin^2\bar\Sigma=\frac{\pi}{2k}\frac{1}{\pi^2/4+\ln^2(k\bar{L})}.\label{sigma_ren_0}
\end{eqnarray}
This result coincides with the general cross-section for a slow particle ($k\bar L\ll1$) interacting with a short-range potential, or even with the scattering in an impenetrable cylinder of radius\footnote{Considering this case, in the limit $r_c\sim\bar L\rightarrow0$, we return to the IR fixed point, $c_-=0$, with a vanishing cross-section.} $r_c=2\bar L/C_E$, see for example pages $551$-$552$ of ref. \cite{Landau_3}. The above formula (\ref{sigma_ren_0}) can also be seen as the cross-section of the renormalized Dirac delta function potential, \cite {Jackiw_Delta,Huang}. Independently of the interpretations, our formula (\ref{sigma_ren_0}) describes a non-trivial isotropic AB scattering which indicates a local short-range interaction.

%%%%%%%%%%%%%%%%%%%%%%%%%%%%%%%%%%%%%%%%%%%%%%%%%%%%%%%%%%%%%%%%%%%%%%%%%%%
\subsection{Physical interpretation for the new scales as bound states}\label{subsec:ligado}

By construction, our regularization procedure  introduces a short-range well potential relevant for  the modes within the interval $0\le\nu_m<1$. This fact  opens the possibility for existence of bound states with certain angular momenta $m=-n$ e $m=-n-1$ that might survive the $R\rightarrow0$ limit. In order to  analyze this option, let us further assume  that $E<0$, and  $k_b^{(m)}=\sqrt{-2ME/\hbar^2}$. The corresponding bound states  wave function needs to be square-integrable at infinity. Thus for the regularized solution for $\zeta\neq0,$  we obtain 
\begin{eqnarray}
u_{m}(r)= \left\{ \begin{array}{ll} 
 D_m\sqrt{r}\,K_{\nu_m}\left(k_b^{(m)}r\right) , \ r>R,\nonumber\\
D_m\sqrt{R}\, K_{\nu_m}\left(k_b^{(m)}R\right)\frac{\sin\left(\rho r\right)}{\sin(\rho R)},  \ \rho=\frac{1}{R}\sqrt{\lambda-(k_b^{(m)}R)^2}, \ r<R,\label{onda_ligada}
\end{array} \right. 
\end{eqnarray}
where $K_{\nu_{m}}(z)$ is the modified Bessel function of the second kind \cite{Bessel}. Once again, by hypothesis $R$ is a small scale such that $k_b^{(m)}R\ll1$, so around $r=R$, $\sin(\rho r)\approx \sin\sqrt{\lambda}$ and we only need the corresponding asymptotic expressions
\begin{eqnarray*}
K_{\nu}(z)\approx \frac{1}{\Gamma(1-\nu)}\left(\frac{z}{2}\right)^{-\nu}-\frac{1}{\Gamma(1+\nu)}\left(\frac{z}{2}\right)^{\nu}, \ z\ll1.
\end{eqnarray*}
Then the continuity of $u'_m(r)/u_m(r)$ at $r=R$ can be read as
\begin{eqnarray}
\sqrt{\lambda}\cot\sqrt{\lambda}-\frac{1}{2}\equiv\gamma=-\nu_{m}\,\frac{1+\frac{\Gamma(1+\nu_m)}{\Gamma(1-\nu_m)}\left(\frac{2}{k_b^{(m)}R}\right)^{2\nu_m}}{1+\frac{\Gamma(1+\nu_m)}{\Gamma(1-\nu_m)}\left(\frac{2}{k_b^{(m)}R}\right)^{2\nu_m}}.
\end{eqnarray}
Now by comparing with equation (\ref{gamma_R}) and solving for $k_b$ we find
\begin{eqnarray}
k_b^{(m)}=\left(\frac{1}{-\varepsilon}\frac{\Gamma(1+\nu_m)}{\Gamma(1-\nu_m)}\right)^{\frac{1}{2\nu_m}}\frac{2}{L_m}.\label{b_state}
\end{eqnarray}
That is, in the case $\varepsilon=1$ there are no bound states at all, while  for $\varepsilon=-1$ each one of the modes $m=-n$ and $m=-n-1$  corresponds to  a bound state when $k_b^{(m)}R\ll1$. In the particular case when the two scales coincide, i.e. for  $L_{-n}=L_{-n-1}\equiv L$, the ratio
%%%%%%%%%%%%FIGURE%%%%%%%%%%%%%%%%%%%%
\begin{figure}[ht]
    \centering
    \includegraphics[scale=0.43]{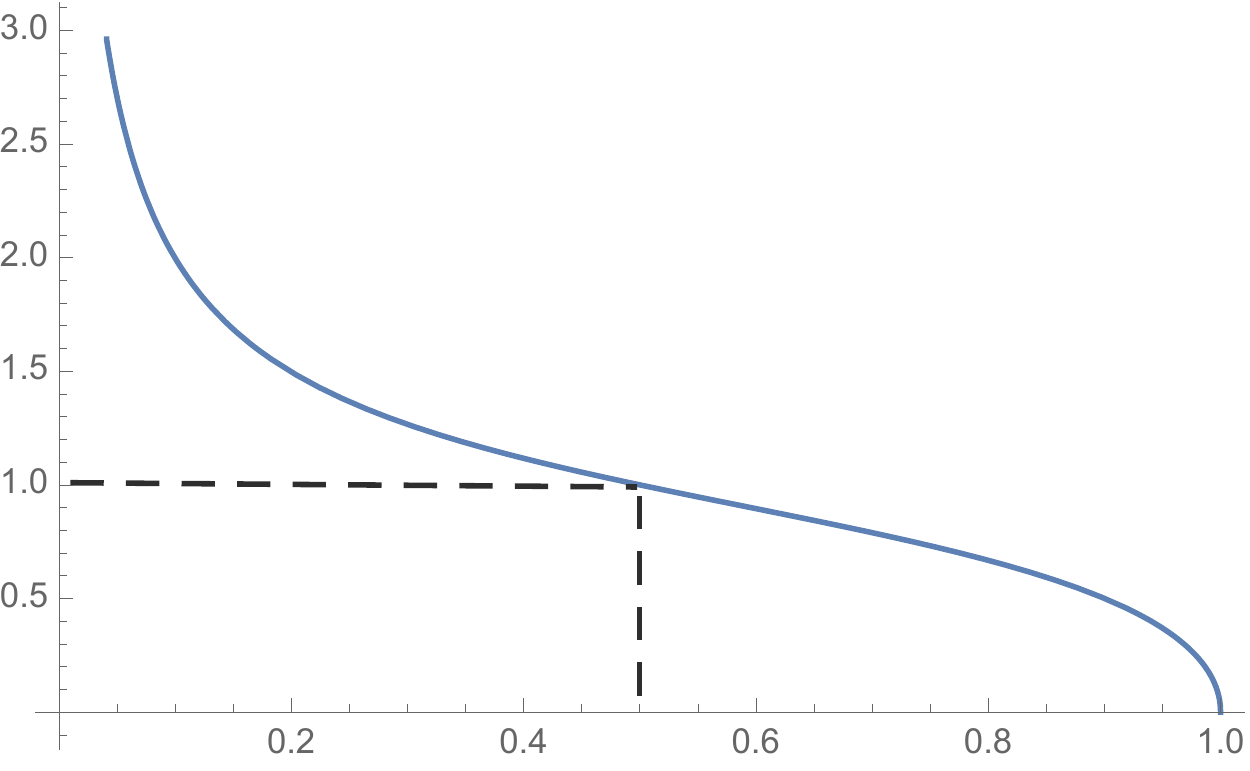}
    \begin{quotation}
    \caption[ed]{\small Equation (\ref{fixada}).}
     \label{fig:energia}
    \end{quotation}
\end{figure}
%%%%%%%%%%%%%%%%%%%%%%%%%%%%%%%%%%%%
\begin{eqnarray}
\frac{k_b^{(-n-1)}}{k_b^{(-n)}}=\left(\frac{\Gamma(1+\zeta)}{\Gamma(1-\zeta)}\right)^{\frac{1}{2\zeta}}\left(\frac{\Gamma(2-\zeta)}{\Gamma(\zeta)}\right)^{-\frac{1}{2(1-\zeta)}}\label{fixada}
\end{eqnarray}
is completely fixed. As one can see from the  curve  $k_b^{(m)} (\zeta)$  shown on figure \ref{fig:energia}, in the range $0<\zeta<1/2$ we have $k_b^{(-n-1)}>k_b^{(-n)}$, while for  $1/2<\zeta<1$ the inverse relation  $k_b^{(-n-1)}<k_b^{(-n)}$ holds; the degeneracy $k_b^{(-n-1)}=k_b^{(-n)}$ occurs for  magnetic fluxes with $\zeta=1/2$. In conclusion, for  $\varepsilon=-1$ the new scales correspond to the energies of certain bound states and we can eliminate them (both $L_{-n}$ and $L_{-n-1}$ as well), in favor of $k_b^{(-n)}$ and  $k_{b}^{(-n-1)}$ in the cross-section. Notice that in the derivation of the bound states energies we have  assumed that  $k_b^{(m)}R\ll1$, which results in $k_b^{(m)}\sim1/L_m$ and  as a consequence we have $R< L_m$, so that such bound states do exist even in the limit $R\rightarrow0$. 

For quantized fluxes $\Phi/\Phi_0\in\mathbb{N}$ (i.e. for $\zeta=0$) the procedure is quite similar. The main differences comes from the behavior of $K_0(z)$ around the origin
\begin{eqnarray*}
K_0(z)\approx -C_E-\ln(z/2), \ z\ll1,
\end{eqnarray*}
and the fact that now $\gamma$ is given by (\ref{gamma_0}). Following the steps we have already made in the previous more general case, we find for the bound state energy 
\begin{eqnarray}
k_b^{(-n)}=\frac{2}{\bar L}.\label{bound_N}
\end{eqnarray}
The cross-section (\ref{sigma_ren_0}) in this particular case has  quite a  clear meaning. Like the two-dimensional scattering in Dirac delta function potential, we are dealing with a non-relativistic exemple of the \emph{dimensional transmutation} due to the existence of a bound state  responsible for the anomalous behavior\footnote{The standard behavior is $\sigma_k\sim 1/k$ by dimensional analysis.}  of the cross-section \cite{Jackiw_Delta,Huang},
\begin{eqnarray}
\bar\sigma_k^{ren}=\frac{\pi}{2k}\frac{1}{\pi^2/4+\ln^2\left(2k/k_b^{(-n)}\right)} \sim\frac{1}{k\ln^2\left(2k/k_b^{(-n)}\right)},  \ k/k_b^{(-n)} \gg1.
\end{eqnarray}

The above discussion about the existence of bound states in the AB problem might sound rather meaningless, since the Calogero potential, $2MV(r)/\hbar^2=\mathcal{C}_m/r^2$, is repulsive in the range $1/2\le\zeta<1$ ($0\le\mathcal{C}_m<3/4$). Thus the bound states can only be associated to the short-range well potentials introduced by the regularization procedure, see figures \ref{fig:VR_rep_at} and \ref{fig:VR_at_at}. But for the specific AB solenoid this makes perfect sense, since  in the interior of a transversal magnetic field the motion of a charged particle is bounded, thus creating the Landau energy levels. So, in the $\varepsilon=-1$ case,  up to two Landau levels --- for which the particle is ``trapped'' within the solenoid --- will survive even in the null radius limit. Indeed, the probability to find   the particle outside the solenoid (in the classically forbidden region) is exponentially suppressed by the behavior $|\sqrt{r}K_{\nu_m}(k_b^{(m)}r)|^2\sim e^{-2k_b^{(m)}r}$ for $k_b^{(m)} r\gg1$. Notice that the two allowed bound states have definite values of the angular momentum in the direction of the field, $m=-n$ and $m=-n-1$, reflecting the fact that the trapped particle  rotates clockwise, due to the negative sign choice, $-|e|$, of its electric charge.

\section{AB  Beta function and RG flows}\label{sec:beta}

The rich variety of options and interesting physical phenomena behind the renormalization procedure needed for  the correct description of the AB scattering (including the ones related to the presence or not  of bound states) represent a subject already  known in the Calogero potential literature, not only for the strong coupling region $\mathcal{C}_m<-1/4$, which is not relevant for the  AB scattering problem, but also in the ``medium-weak''-range $-1/4\le\mathcal{C}_m<3/4$, see \cite{medium_AB}. On the other hand, it is not usual (although there are some exceptions \cite{beta1,beta2}) to explore how all these options can be understood in a clear and  elegant way within the framework of  the renormalization group (RG) methods and, related to it, to the concepts of critical points, phase transitions and RG flows.  This section is dedicated to this subject.

\begin{figure}[ht]
    \centering
    \includegraphics[scale=0.3]{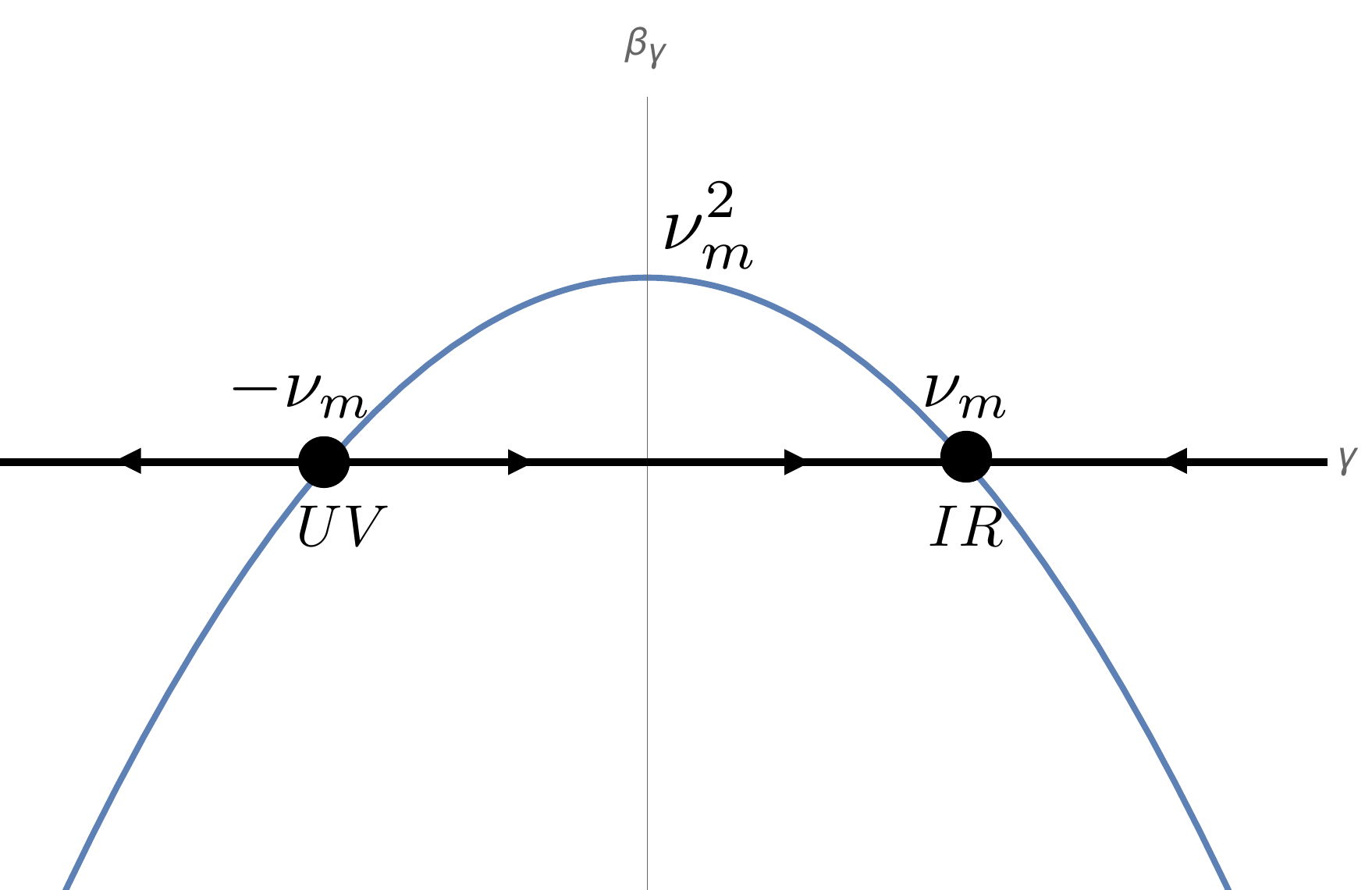}
    \begin{quotation}
    \caption[ed]{\small Beta function (\ref{beta}). The arrows indicate the RG flow and the dots are the fixed points.}
     \label{fig:beta}
    \end{quotation}
\end{figure}

The ``running coupling'' $\gamma $ for each $m$-mode is given by (\ref{gamma_R}) (for $\zeta\neq0$ and $m=-n,-n-1$) or by  (\ref{gamma_0}) (for $\zeta=0$ and $m=-n$). Defining the adimensional scale $l=\ln(R/L_m)$ and taking the derivative of (\ref{gamma_R}) one can easily obtain the exact beta function for the considered AB scattering problem
\begin{eqnarray}
\frac{d\gamma}{dl}\equiv\beta_{\gamma}=-\left(\gamma+\nu_m\right)\left(\gamma-\nu_m\right),\label{beta}
\end{eqnarray}
which turns out to have two fixed (critical) points $\beta_{\gamma} (\gamma^*)=0$. The IR stable one corresponds to  $\gamma_{IR}=\nu_m$ ($c_-=0$ in (\ref{U0})) and a conformal dimension $\Delta_{IR}=-2\nu_m$, while the  unstable UV fixed point is characterized by  $\gamma_{UV}=-\nu_{m}$ ($c_+=0$ in (\ref{U0})) and a conformal dimension $\Delta_{UV}=2\nu_m$. The beta function behavior, together with the RG flow relating these two critical points, is shown on figure \ref{fig:beta}. By construction, Eq. (\ref{gamma_R}) is the general solution of (\ref{beta}) with  $\varepsilon L_m^{2\nu_m}$ as an integration constant. Let us rewrite the equation in terms of the new parameter $l$
\begin{eqnarray}
\gamma(l)=\nu_m\frac{1-\varepsilon e^{2\nu_ml}}{1+\varepsilon e^{2\nu_ml}}.\label{gamma_l}
\end{eqnarray}
Now we will analyze the role of the $\varepsilon$ sign. 
%%%%%%%%%%%%%%%%%%%%%%%%%%%%%%%%%%%
%%%%%%%%%%%%%%%%%%%%%%%%%%%%%%%%%%%%

\begin{itemize}

\item[i) $\varepsilon=1$:]

If $\varepsilon=1$, then $l$ runs over all the scales, $-\infty<l<\infty$, and  $\gamma_{UV}=-\nu_m<\gamma<\nu_m=\gamma_{IR}$, as shown by the blue curve in figure \ref{fig:gamma_l}. This RG flow doesn't have a characteristic length scale (in the RG language it is a \emph{massless} flow) going from the UV fixed point to the IR one. As one can see from figure \ref{fig:gamma_la} the original coupling $\lambda \in (\lambda_{IR},\lambda_{UV})$, with $\lambda_{IR}<0$ and $\lambda_{UV}>0$ when $\nu_m>1/2$. Therefore a bound state is impossible as $\lambda$ can change sign in the vicinities of the IR point and the well potential turns into a repulsive barrier potential. The absence of natural scales prevents the existence of bound states.

Note that our  specific choice $-\infty<\lambda<\pi^2$ ($-\infty<\gamma<\infty$) of the $\lambda$-interval  is crucial for the compatibility among the RG flow description and the regularized potential --- this fact  was missed in ref. \cite{Beane,medium_AB}, where $\lambda>0$ is  assumed. As the cutoff is unbounded, for large $R$ a bound state would be unavoidable if $\lambda$ was always positive, thus leading to an inconsistency. Already   in the vicinities of the UV energy regime the condition $kR\ll1$ still holds, and in the regularized region $r<R$ we have a deep well potential ($\sim\lambda/R^2$), which does not support a bound state and diverges in the $R\rightarrow0$ and  $\lambda\rightarrow\lambda_{UV}^-$ limits.

\vspace{3mm}

\item[ii) $\varepsilon=-1$:]
The conclusions are absolutely different for the $\varepsilon=-1$ case. Now the scale $R$ has $L$ as its maximum or minimum value (in the RG language this corresponds to a  massive flow) since $\gamma(l)$ diverges when $l\rightarrow0$. For the  ``initial scale condition'' $l_0<0$ ($R_0<L_m$) we have $-\infty<\gamma(l)<-\nu_m$ with $-\infty<l<0$, see the red line on the third quadrant of figure \ref{fig:gamma_l}, so that  $L_m$ is the maximum value permitted  for the cut-off $R$. Therefore, along  the entire RG flow $\gamma<1/2$ ($\lambda>0$) the ``short-range'' potential is represented by the potential well. In this case, the $R$-independent bound state given by Eq. (\ref{b_state}) is present  even if the Calogero potential is repulsive, because (\ref{b_state}) implies that $k_b^{(m)}\sim 1/L_m$ and thus the existence condition for such a state, $k_b^{(m)}R\sim R/L_m\ll1$, is fulfilled in the $R\rightarrow0$, $\lambda\rightarrow\lambda_{UV}^+$ limits. Now the meaning of $\lambda_{UV}$ becomes evident: it is the critical coupling value that regulates the existence ($\lambda>\lambda_{UV}$) or not ($\lambda<\lambda_{UV}$) of the bound state in the $R\rightarrow0$ limit.
%%%%%%%%%%%%FIGURE%%%%%%%%%%%%%%%%%%%%
\begin{figure}[ht]
    \centering
    \includegraphics[scale=0.3]{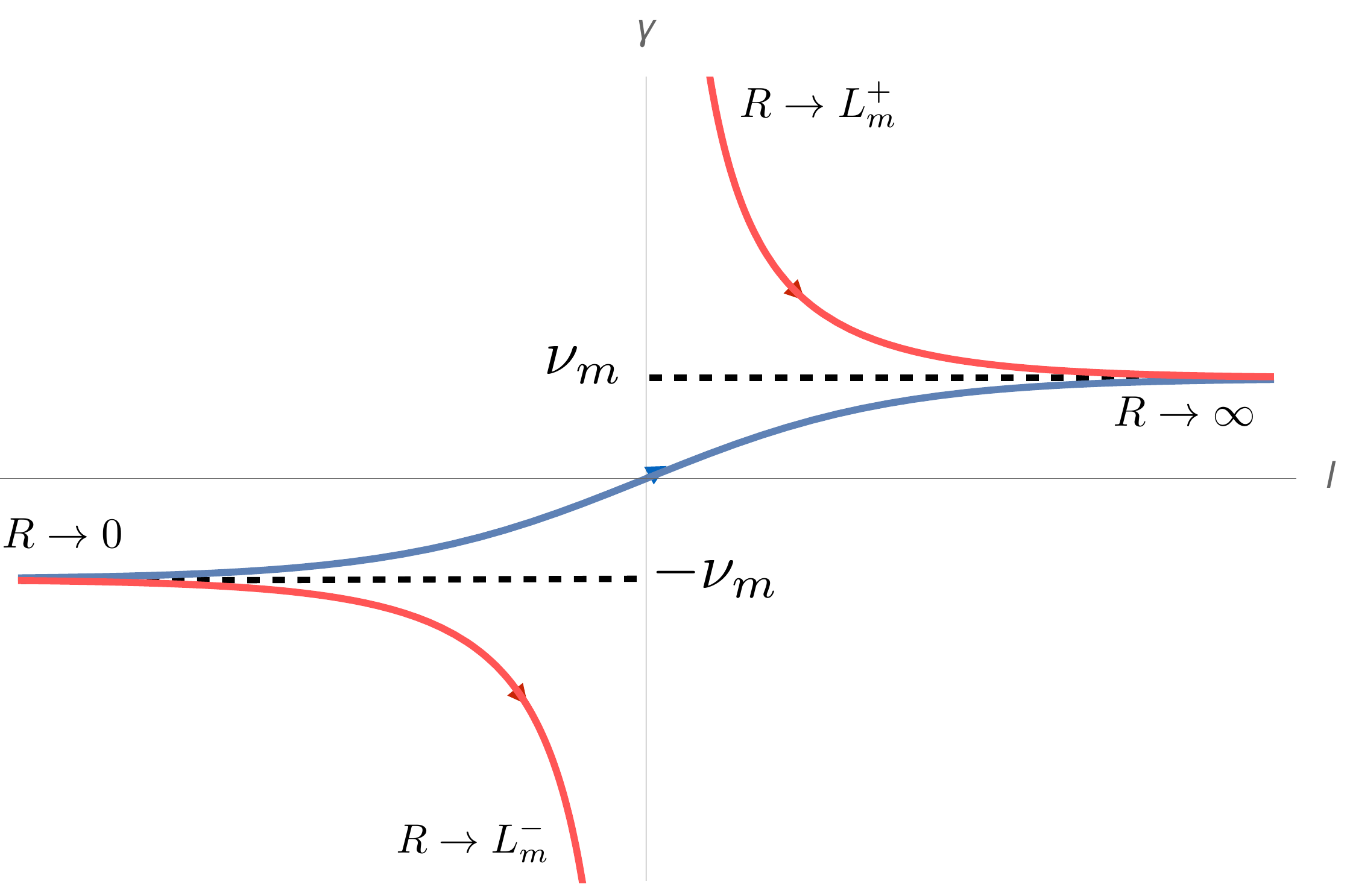}
    \begin{quotation}
    \caption[ed]{\small Curves for Eq. (\ref{gamma_l}); the blue one is for $\varepsilon=1$ and the red ones are for the $\varepsilon=-1$ case.}
     \label{fig:gamma_l}
    \end{quotation}
\end{figure}
%%%%%%%%%%%%%%%%%%%%%%%%%%%%%%%%%%%%

Another possibility for $\varepsilon=-1$ occurs when the  RG ``initial condition'' is given by  $\gamma(l_0)>\nu_m$. The running coupling range is now $\nu_m<\gamma<\infty$ and $l>0$, see  red curve on the first quadrant of  figure \ref{fig:gamma_l}. Since $\gamma$ can be arbitrarily large ($R\rightarrow L^+$) and consequently $|\lambda|\gg1$, with $\lambda<0$, so that  in this limit the potential in the region $r<R$ is a repulsive barrier.  Hence in this case we still have a characteristic length, but not a bound state. In fact, the condition $k_b^{(m)}R\sim R/L_m\ll1$ cannot be satisfied. Such a  flow is not related with the Calogero potential renormalization, since the limit $R\rightarrow0$ is now not available. In fact, it represents the ``inverse situation'' --- a physical barrier potential regularized by an IR Calogero cutoff ($R>L_m$), and  at the end of  the renormalization the limit $R\rightarrow\infty$ has to be taken. Thus it will be excluded from our analysis.
\end{itemize}

\vspace{0.5cm}

When the magnetic flux is exactly $\Phi=N\Phi_0$, $N\in\mathbb{N}^*$, the UV and IR fixed points of the mode $m=-n$ ($\nu_{-n}=0$) ``collide'' in one new marginal point, $\gamma_{UV}=\gamma_{IR}=0$ and the maximum of the parabola in fig. \ref{fig:beta} touches the horizontal axis. The beta function for this case is
\begin{eqnarray}
\beta_{\gamma}=-\gamma^2.\label{beta_2}
\end{eqnarray}
The passage from (\ref{beta}) to (\ref{beta_2}) is known as a BKT-like phase transition \cite{B1_BKT,B2_BKT,KT_BKT}. The integration of (\ref{beta_2}) leads to (\ref{gamma_0}) with $\bar c$ being the integration constant. In this limit the \emph{massless} flow between the two fixed points disappears from the figures \ref{fig:beta} and \ref{fig:gamma_l}. Now only the $\gamma=0$ fixed point has no intrinsic scale and the conformal symmetry takes place for  the solution with $\bar c=0$. To its left ($\bar c<0$) we have an unstable massive flow with a maximal scale $L_{-n}$, and to its right ($\bar c>0$) the massive stable flow with  a minimal scale $L_{-n}$. Like in the previous case, the bound state is associated to the unstable flow, because the limit $R\rightarrow0$ can be taken. The stable massive case must be excluded once again, since  the region $r<R$ cannot be made arbitrary small. The absence of parameters in the beta function (\ref{beta_2}) show us that all the magnetic fluxes $N\Phi_0$ belong to the same universality class, reflecting the periodic behavior of the physical observables in the AB Scattering. It also  suggests that the $\bar L$-scale (defining the bound state  energy (\ref{bound_N})) is independent of $N$ as well.

\vspace{0.5cm}

Our last comment is about the attractive region of the Calogero potential that is not realized in the AB Scattering, i.e.  the case of  ``strong-attraction''-coupling  $\mathcal{C}_m<-1/4$. We can adapt our results for this case as well  with a formal change of the variables: $\nu_m\rightarrow ig$ ($g\in\mathbb{R}$). Now the beta function takes the form 
\begin{equation}
\beta_g(\gamma)=-\gamma^2-g^2,
\end{equation}
without any real zeros; that is, the maximum of the parabola on  fig. \ref{fig:beta} is now below the horizontal axis. By integrating this beta function one can find  the running-coupling scale dependence   $\gamma(l)=-g\tan(gl-\phi_0)$. Its  $2\pi/g$-periodic behavior, which corresponds to the periodic region $\lambda>\pi^2$ in the definition of  $\gamma(\lambda)=\sqrt{\lambda}\cot\sqrt{\lambda}$, generates an infinite chain of scales, leading to an infinite number of bound states accumulated in the low energies region. This phenomenon is known as  the \emph{Efimov effect} commented earlier in section \ref{sec:ren}.

\section{Conclusions}\label{sec:con}

In this paper we have revisited in details the elastic scattering of a non-relativistic charged particle without spin in an ideal magnetic field $\vec B=B_0\,\delta(x)\delta(y)\hat z$ --- the AB scattering problem. Since  our analysis  is based on  the phase-shifts method,  we have rewritten the incident wave function $e^{ikx}e^{i\alpha(\varphi-\pi)}$ as $e^{ikx}+2ie^{ikx}e^{i\alpha(\varphi-\pi)/2}\sin[\alpha(\varphi-\pi)/2]$ and the second term was ``hidden'' in the outgoing cylindrical wave function. Our  choice of  standard scattering boundary conditions allows us to derive  the AB scattering amplitude in an easy and direct way, including an extra $\delta$-term, important for the unitarity of the S-Matrix.

We have implemented, in our study of the  correct  self-adjoint extension  of the cylindrical  AB  Hamiltonian, the renormalization methods specific for the one-dimensional Calogero model, by taking into account the advantages of the established  AB scattering/Calogero correspondence. We have demonstrated that, depending on the values of the magnetic flux,  one or  two of the phase-shifts modes have to be renormalized. As a consequence, the  AB cross-section receives certain renormalization corrections, parametrized by at most two new scales  associated to the bound states energies. We expect  that our approach, combining the phase-shift method and the renormalization group, could be successfully  adapted to the correct description of the AB scattering of non-relativistic spin-$1/2$ particles, by considering an appropriate self-adjoint extension of the Pauli equation, as well as to  the relativistic RG corrections to the high-energy electron scattering cross-section in terms of the renormalized solutions of  the  Dirac equation in an external magnetic field created by  an infinitely thin solenoid.

\vspace{0.5cm}

Since the emergence of new physical scales appears to be one of the essencial elements of our renormalization of the AB scattering amplitudes, we find worthwhile  to comment  once more  on their physical  meaning. One can adopt a rather conservative  point of view, observing that the finite solenoid problem has also two natural scales:  the solenoid radius $r_c$ and the magnetic field strength  $r_B=\left(B|e|/\hbar\right)^{-1/2}$. In the ``infinitely thin solenoid'' limit, both scales go to zero with a fixed ratio $\left(r_c/r_B\right)^2=2\Phi/\Phi_0$. Therefore the physical results are  functions of this ratio only, and the conformal symmetry should be preserved. One can further try to relate the renormalization scales with these natural scales, e.g. $r_c\sim L_{-n}$ and $r_B\sim L_{-n-1}$, so the renormalized cross-section we have derived can be regardaded as describing a more realistic scattering problem involving a finite solenoid and a finite magnetic field. Apparently this does not seems to be the case,  since after renormalization the $R$ cutoff becomes arbitrarily small, making clear that we are dealing with an infinitely thin solenoid, and also because of the fact that our cross-section does not agree with the results obtained in  the WKB approximation for solenoids of finite radius \cite{cilindro_AB2}.

An argument in favor of the quantum origin of  the new energy scales  comes from the fact that all the different scattering amplitude corrections predicted by the renormalization process can be explained by the RG flow of the $\gamma$-running coupling. We have demonstrated that a few different RG flows (massive and massless) can be realized. The case $\varepsilon=1$ represents a massless RG flow connecting  the UV and IR regimes without any characteristic length scale. Instead, for the $\varepsilon=-1$ case a massive RG flow occurs and now the scale $L_m$ appears as the maximum or the minimum value of the $R$ cutoff. Since our interest is in short-range values of the cutoff, it is assumed that $L_m$ is the greatest length scale corresponding to the flow that leaves the UV fixed point and goes to minus infinity ($-\infty<\gamma<-\nu_m$). Along this flow, in the region $r<R$, there is a potential well that  supports one bound state with energy $\sim1/L_m^2$. As we have shown, one or two of the Landau levels can survive and the cross-section appears to be  a function of the corresponding bound states energies. This phenomenon provides an example of dimensional transmutation, where the conformal symmetry is spontaneously broken during the renormalization process that creates certain quantum scale anomaly. Their exact values can be determined by experimental observation only. It is important to emphasize that, in all the cases, the IR fixed point is a part of the coupling-space. So we can have two, one or zero new scales; in the last case the conformal symmetry is still preserved and the AB cross-section remains unchanged coinciding with the original one (\ref{s_AB1}).

\vspace{0,5cm}

The most relevant aspect arising from the renormalization of the AB Scattering problem is the universal phase-shift deviation $2\delta_{-N}^{ren}+N\pi=2\bar\Sigma(k)\neq0$ for the quantized magnetic flux case, $\Phi=N\Phi_0$ (e.g. when the cylinder is a superconductor), leading to a non null, isotropic and energy dependent cross-section, and its relation with a BKT-like phase transition of infinite order. This is an explicit realization of the phase transition in the Calogero potential beyond the Efimov effect --- another recent case was observed in a graphene plane at zero temperature \cite{graphene}.

\begin{flushleft}
\textbf{Acknowledgments}

I wish to thank E. S. Costa, G. Luchini and G. Sotkov for the productive discussions and G. Sotkov for the great help in the elaboration process of this manuscript.
\end{flushleft}

\appendix
\section{Phase-shifts method for short-range two-dimensional elastic scattering}\label{sec:esp_2d}

In this appendix we present a brief review of the quantum elastic scattering in a plane to support the analysis made in the main text. The wave function that describes a particle with energy $E$ launched from $x\rightarrow-\infty$ with momentum $\vec p_{in}=\hbar k\,\hat x =\sqrt{2ME}\,\hat x$ and scattered by a short range external potential obeys the boundary condition
\begin{eqnarray}
\psi(r,\varphi)\approx e^{ikx}+f_k(\varphi)\frac{e^{ikr}}{\sqrt{-ir}}, \ x=r\cos\varphi,\  kr\gg1.\label{1}
\end{eqnarray}
The angular function $f_k(\varphi)$ is the scattering amplitude and the cross-section is defined by $\sigma_k=|f_k(\varphi)|^2$. This asymptotic behavior is fulfilled by the following \emph{ansatz}, see pages $507-508$ of \cite{Landau_3} 
\begin{equation}
\psi_k(r,\varphi)=\sum_{m=-\infty}^{\infty}i^me^{i\delta_m}Q_m^{(k)}(r;\delta_m)e^{im\varphi},\label{3}
\end{equation}
where the function $Q_m^{(k)}(r;\delta_m)$ is the solution of the cylindrical equation such that
\begin{equation}
Q_m^{(k)}(r;\delta_m)\approx\sqrt{\frac{2}{kr}}\sin\left(kr-\frac{\pi}{2}\left(m-\frac{1}{2}\right)+\delta_m\right), \ kr\gg1.\label{2}
\end{equation}
The quantities $\delta_m$, $m\in\mathbb{Z}$, are the phase-shifts which completely describe the scattering process. Substituting the asymptotic expression (\ref{2}) in (\ref{3}) and after some algebraic manipulations making use of the identity, see page $114$ of \cite{Landau_3}
\begin{equation}
e^{ikx}=\sum_{m=-\infty}^{\infty}i^mJ_{m}(kr)e^{im\varphi},\label{livre}
\end{equation}
the result (\ref{1}) is easily found with the scattering amplitude being given by
\begin{equation}
f_k(\varphi)=\frac{1}{i\sqrt{2\pi k}}\sum_{m=-\infty}^{\infty}\left(e^{2i\delta_m}-1\right)e^{im\varphi}.\label{f_k}
\end{equation}
So the elastic scattering problem can be summarized in the search of the phases $\delta_m$ ($m\in\mathbb{Z}$) defined by equation (\ref{2}), together with the summation (\ref{f_k}). Being not necessary the direct use of the total wave function (\ref{3}).

\section*{References}

\bibliography{mybibfile}

\begin{thebibliography}{10}
\expandafter\ifx\csname url\endcsname\relax
  \def\url#1{\texttt{#1}}\fi
\expandafter\ifx\csname urlprefix\endcsname\relax\def\urlprefix{URL }\fi
\expandafter\ifx\csname href\endcsname\relax
  \def\href#1#2{#2} \def\path#1{#1}\fi

\bibitem{Peshkin}
M.~Peshkin, A.~Tonomura, The Aharonov-Bohm Effect, Lecture Notes in Physics,
  Springer; Softcover reprint of the original 1st ed. 1989 edition, 2013.

\bibitem{Tonomura_new}
A.~Tonomura, The ab effect and its expanding applications, Journal of Physics
  A: Mathematical and Theoretical 43~(35).

\bibitem{AB_original}
Y.~Aharonov, D.~Bohm, Significance of electromagnetic potentials in the quantum
  theory, Phys. Rev. 115~(3).

\bibitem{experimento_1_AB}
R.~G. Chambers, Shift of an electron interference pattern by enclosed magnetic
  flux, Phys. Rev. Lett. 5~(3).

\bibitem{Wu_Yang}
T.~T. Wu, C.~N. Yang, Concept of nonintegrable phase factors and global
  formulation of gauge fields, Phys. Rev. D 12~(12) (1975) 3845--3857.

\bibitem{Merzbacher}
E.~Merzbacher, Single valuedness of wave functions, American Journal of Physics
  30~(237).

\bibitem{Ruji_AB}
S.~Ruijsenaars, The aharonov-bohm effect and scattering theory, Annals of
  Physics 146~(1) (1983) 1 -- 34.

\bibitem{Gitman}
D.~Gitman, I.~Tyutin, B.~Voronov, Self-adjoint Extensions in Quantum Mechanics:
  General Theory and Applications to Schr\"odinger and Dirac Equations with
  Singular Potentials, 1st Edition, Birkh\"auser, 2012.

\bibitem{Beane}
S.~R. Beane, P.~F. Bedaque, L.~Childress, A.~Kryjevski, J.~McGuire, U.~V.
  Kolck, Singular potentials and limit cycles, Phys. Rev. A 64 (2001) 042103.

\bibitem{Gitman2}
D.~Gitman, I.~Tyutin, B.~Voronov, Self-adjoint extensions and spectral analysis
  in calogero problem, Journal of Physics A: Mathematical and Theoretical
  43~(14) (2010) 145205.

\bibitem{Cal1}
F.~Calogero, Solution of a three‐body problem in one dimension, Journal of
  Mathematical Physics 10.

\bibitem{Cal3}
F.~Calogero, Solution of the one‐dimensional n‐body problems with quadratic
  and/or inversely quadratic pair potentials, Journal of Mathematical Physics
  12.

\bibitem{Wein_Col}
S.~Coleman, E.~Weinberg, Radiative corrections as the origin of spontaneous
  symmetry breaking, Phys. Rev. D 7 (1973) 1888--1910.

\bibitem{B1_BKT}
V.~Berezinskii, Destruction of long-range order in one-dimensional and
  two-dimensional systems having a continuous symmetry group i. classical
  systems, Sov. Phys. JETP 32~(3) (1970) 493.

\bibitem{B2_BKT}
V.~Berezinskii, Destruction of long-range order in one-dimensional and
  two-dimensional systems possessing a continuous symmetry group. ii. quantum
  systems, Sov. Phys. JETP 34~(3) (1971) 610.

\bibitem{KT_BKT}
J.~M. Kosterlitz, D.~J. Thouless, Ordering, metastability and phase transitions
  in two-dimensional systems, Journal of Physics C: Solid State Physics 6~(7)
  (1973) 1181.

\bibitem{Bessel}
A.~Erdelyi, H.~Bateman, Higher Transcendental Functions Bateman Manuscript
  Project Volume I One 1, 1st Edition, McGraw-Hill Book Company, Inc., 1953.

\bibitem{berry_eleg}
M.~V. Berry, Aharonov-bohm wavefunction obtained by applying dirac's magnetic
  phase factor, Eur. J. Phys. 1 (1980) 220--144.

\bibitem{Sakoda_AB}
S.~Sakoda, M.~Omote, Aharonov-mohm scattering: The role of the incident wave,
  Journal of Mathematical Physics 38~(2).

\bibitem{espalhamento_AB_WKB}
Y.~Sitenko, N.~Vlasii, The aharonov--bohm effect in scattering theory, Annals
  of Physics 339 (2013) 542 -- 559.

\bibitem{cilindro_AB1}
Y.~A. Sitenko, The aharonov-bohm effect in scattering of nonrelativistic
  electrons by a penetrable magnetic vortex, Quantum Studies: Mathematics and
  Foundations 1~(3 - 4) (2014) 213 -- 222.

\bibitem{cilindro_AB2}
O.~Yilmaz, Scattering of a charged particle from a hard cylindrical solenoid:
  Aharonov-bohm effect, Chin.J.Phys. 52.

\bibitem{Hagen_AB}
C.~R. Hagen, Aharonov-bohm scattering amplitude, Physical Review D 41~(6).

\bibitem{Cal_s.a.}
J.~Derezi\'nski, S.~Richard, On schr\"odinger operators with inverse square
  potentials, Ann. Henri Poincar\'e 18~(3) (2017) 869 -- 928.

\bibitem{Efimov_rev}
P.~Naidon, S.~Endo, Efimov physics: a review, Reports on Progress in Physics
  80~(5) (2017) 056001.

\bibitem{Efimov}
V.~N. Efimov, Weakly-bound states of three resonantly-interacting particles,
  Soviet Journal of Nuclear Physics 12~(5) (1970) 589--591.

\bibitem{Efimov2}
V.~Efimov, Energy levels arising from resonant two-body forces in a three-body
  system, Physics Letters B 33~(8) (1970) 563 -- 564.

\bibitem{Ren_1}
H.~E. Camblong, L.~N. Epele, H.~Fanchiotti, C.~A.~G. Canal, Renormalization of
  the inverse square potential, Physical Review Letters 85~(8).

\bibitem{Arg1}
H.~E. Camblong, L.~N. Epele, H.~Fanchiotti, C.~A.~G. Canal, Dimensional
  transmutation and dimensional regularization in quantum mechanics: I. general
  theory, Annals of Physics 287~(1) (2001) 14 -- 56.

\bibitem{Ren_2}
S.~A. Coon, Anomalies in quantum mechanics: the 1/r2 potential, American
  Journal of Physics 70~(513).

\bibitem{Ren_3}
E.~Braaten, D.~Phillips, The renormalization group limit cycle for the 1/r2
  potential, Physical Review A 70~(5) (2004) 052111.

\bibitem{Ren_4}
H.~E. Camblong, L.~N. Epele, H.~Fanchiotti, C.~A.~G. Canal, C.~R.
  Ord{\'o}{\~n}ez, On the inequivalence of renormalization and self-adjoint
  extensions for quantum singular interactions, Physics Letters A 364~(6)
  (2007) 458 -- 464.

\bibitem{Ren_5}
C.~Burgess, P.~Hayman, M.~Williams, L.~Zalav{\'a}ri, Point-particle effective
  field theory i: classical renormalization and the inverse-square potential,
  Journal of High Energy Physics 2017~(4) (2017) 106.

\bibitem{medium_AB}
D.~Bouaziz, M.~Bawin, Singular inverse-square potential: Renormalization and
  self-adjoint extensions for medium to weak coupling, Phys. Rev. A 89~(2)
  (2014) 022113.

\bibitem{Landau_3}
L.~D. Landau, L.~M. Lifshitz, Quantum Mechanics, Course of Theoretical Physics
  Vol. 3, 3rd ed. (Elsevier Butterworth-Heinemann), 1977.

\bibitem{Jackiw_Delta}
R.~Jackiw, Diverse Topics in Theoretical and Mathematical Physics, World
  Scientific Publishing Company, 1995.

\bibitem{Huang}
K.~Huang, Quarks, Leptons and Gauge Fields, 2nd Edition, World Scientific
  Publishing Company, 1992.

\bibitem{beta1}
D.~B. Kaplan, J.-W. Lee, D.~T. Son, M.~A. Stephanov, Conformality lost, Phys.
  Rev. D 80~(12) (2009) 125005.

\bibitem{beta2}
K.~M. Bulycheva, A.~S. Gorsky, Limit cycles in renormalization group dynamics,
  Physics-Uspekhi 57~(2) (2014) 171.

\bibitem{graphene}
O.~Ovdat, J.~Mao, Y.~Jiang, E.~Y. Andrei, E.~Akkermans, Observing a scale
  anomaly and a universal quantum phase transition in graphene, Nature
  Communications 8~(1) (2017) 507.

\end{thebibliography}

\end{document}